\documentclass[a4paper,10pt]{article}
\usepackage{amsmath}

\usepackage[small]{titlesec}
\titleformat*{\section}{\bfseries}
\titleformat{\subsection}[runin]{\bfseries\itshape}{\thesubsection.}{3pt}{\space}[.]
\titleformat{\subsubsection}[runin]{\itshape}{\thesubsection.}{3pt}{\space}[.]

\usepackage[labelfont=sc]{caption}

\usepackage[english]{babel}
\usepackage{color}
\usepackage{amsmath, latexsym, amsthm, amsfonts,bm,amssymb,mathscinet} 
\usepackage{natbib}
\usepackage[text={5.6in,8.1in},centering]{geometry}
\usepackage{epsfig}

\usepackage[utf8]{inputenc}
\usepackage[colorlinks,linkcolor=blue, citecolor=blue,urlcolor=blue]{hyperref}
\usepackage[dvipsnames]{xcolor}
\usepackage{enumitem}
\usepackage{bbm}
\usepackage[normalem]{ulem}

\usepackage{booktabs} \newcommand{\tablestretch}[1]{\renewcommand{\arraystretch}{#1}}
\usepackage{media9}
\usepackage[buttonsize=0.7em]{animate}
\usepackage[short]{optional} %still long short (for arxiv)

\newcommand{\dd}{{\,\mathrm d}}
\newcommand{\si}{\sigma}

\renewcommand{\th}{\theta}

\newcommand{\eps}{\varepsilon}

\renewcommand{\phi}{\varphi}

\renewcommand{\scr}[1]{{\mathcal #1}}

\newcommand{\EE}{\mathbb{E}}

\newcommand{\PP}{\mathbb{P}}
\newcommand{\ind}{\mathbf{1}}
\newcommand{\RR}{\mathbb{R}}

\newcommand{\bem}{\begin{bmatrix}}
\newcommand{\enm}{\end{bmatrix}}
\newcommand{\bs}[1]{{\boldsymbol #1}}
\newcommand{\T}{{\prime}}
\newcommand{\diag}{\operatorname{diag}}

\providecommand{\trace}{{\operatorname{tr}}}
\renewcommand{\P}{\mathbb{P}} %overwriting P = paragraph
\newcommand{\GP}{\ensuremath{\mathcal{G}\mathcal{P}}}
\theoremstyle{definition}
\newtheorem{thm}{Theorem}[section]
\newtheorem{prop}[thm]{Proposition}

\newtheorem{rem}[thm]{Remark}
\newtheorem{ex}[thm]{Example}
\newtheorem{defn}[thm]{Definition}
\newtheorem{ass}[thm]{Assumption}
\newtheorem{alg}[thm]{Algorithm}

\newcommand{\expec}[1]{\ensuremath{\mathbb{E}\mspace{-1mu}\left[#1\right]}}

\newcommand{\simind}{\stackrel{\rm ind}{\sim}}

\newcommand{\Bm}{\begin{bmatrix}}
\newcommand{\Em}{\end{bmatrix}}
\newcommand{\Hd}{P}
\newcommand{\Md}{M^\dagger}

\DeclareUnicodeCharacter{03A3}{\Sigma}
\newcommand{\N}{\mathrm{N}}
\title{\bf Continuous-discrete smoothing of diffusions}

\author{\bf  \small Marcin Mider\thanks{Hohenlindener Str. 1, 81677 München, Germany}\\
\small Trium Analysis Online GmbH\\
\small Email: marcin.mider@gmail.com
\and \small \bf Moritz Schauer\thanks{ Chalmers Tvärgata 3, 41296 Göteborg, Sweden}\\
\small Chalmers University of Technology\\\small and University of Gothenburg\\
\small E-mail: smoritz@chalmers.se\\
\and \small \bf Frank van der Meulen\thanks{Mekelweg 4, 2628 CD Delft, The Netherlands }\\
\small Delft Institute of Applied Mathematics (DIAM) \\
\small Delft University of Technology\\
\small Email: f.h.vandermeulen@tudelft.nl
}

\begin{document}
\maketitle

%\tableofcontents

\begin{abstract} 
\parindent0pt
\parskip1ex

Suppose $X$ is a multivariate diffusion process that is observed discretely in time. At each observation time, a  transformation of the state of the process is observed with noise. The smoothing problem consists of recovering the path of the process, consistent with  the observations. We derive a novel Markov Chain Monte Carlo algorithm to sample from the exact smoothing distribution. The resulting algorithm is called the {\it Backward Filtering Forward Guiding (BFFG) algorithm}. We extend the algorithm to include parameter estimation.  The proposed method relies on guided proposals introduced in \cite{Schauer}.  We illustrate its efficiency in a number of challenging problems.

\sloppy 
\noindent
 \emph{{Keywords:} chemical reaction network, data assimilation, diffusion bridge, filtering, guided proposal, Lorenz system, Markov Chain Monte Carlo, partial observations, stochastic heat equation on a graph}.  
 
\noindent \textbf{MSC 2010 subject classifications:} Primary 60J60, 65C05; secondary 62F15.
 \end{abstract}

%\tableofcontents

\numberwithin{equation}{section}
%------
\section{Introduction}
Suppose $X$ is a diffusion process with dynamics governed by the stochastic differential equation (SDE)
\begin{equation}\label{eq:sde} \dd X_t = b(t,X_t) \dd t + \si(t,X_t) \dd W_t \end{equation}
 with prescribed initial density $X_{t_0} \sim \pi$.
Here  $b\colon\, \RR_+\times \RR^d \to \RR^d$ and  $\si\colon\, \RR_+\times \RR^d \to \RR^{d\times d'}$ are the drift and dispersion coefficient respectively.    The process $W$ is a vector valued process in $\RR^{d'}$ consisting of independent Brownian motions. It is assumed that the required conditions for existence of a strong solution are satisfied (cf.\ \cite{Karatzas-Shreve}).  We assume  observation times  $0=t_0 < t_1<\cdots < t_n$ and  observations 
\begin{equation}\label{eq:obs-scheme}	V_i \mid X_{t_i} \sim k_i( X_{t_i}; \cdot ), \qquad i=0,\ldots, n,  \end{equation} with conditional density $k_i$. This includes as special case
\begin{equation}\label{eq:lin-obs}
V_i \mid X_{t_i} \sim \phi(\cdot;  L_i X_{t_i}, \Sigma_i),
\end{equation}  where $L_i$ is a  $m_i \times d$-matrix,  $\Sigma_i$ an $m_i \times m_i$ positive definite matrix and  $\phi(x; \mu,\Sigma)$ denotes the density of the $\operatorname{N}(\mu, \Sigma)$-distribution, evaluated at $x$.    
For $t\ge 0$, denote the set of non-past observations by $\scr{V}_t$, i.e.\ $\scr{V}_t =  \{V_i \colon t_i\ge t\}$. With slight abuse of notation, set $\scr{V}_i= \scr{V}_{t_i}$.

In this article we address two related problems: {\it smoothing} and {\it inference}. For the first one we assume that $b$, $\si$ and $\{k_i, \, i=0,\ldots, n\}$ are known and  aim to reconstruct the path $\{X_t,\, t\in [0,t_n]\}$ based on $\scr{V}_0$, which here is given precise meaning as sampling from the conditional distribution of $X$ on $[0,t_n]$ conditional on $\scr{V}_0$.  Because we are dealing with ``continuous dynamics-discrete observations'', this setup is also referred to as continuous-discrete smoothing or hybrid smoothing. This is an interesting problem in its own right that arises naturally for instance in statistical  analysis of tracking systems (cf.\ \cite{Sarkka-book}). 

Additionally, smoothing constitutes a central component of Bayesian inference algorithms for discretely observed diffusions.  If in \eqref{eq:sde} the coefficients of the diffusion $X$ or the observation scheme $\{k_i,\, i=0,\ldots,n\}$ depend on some unknown parameters, inference  can be performed by employing a data augmentation algorithm, where one iterates between {\it (i)} updating the parameter conditional on the smoothed path and {\it (ii)} continuous-discrete smoothing conditional on the value of the parameter (cf.\ \cite{RobertsStramer}). 
The relevance of smoothing and parameter estimation for diffusions is illustrated from the problem reappearing in a wide variety of application fields, see for instance \cite{VanKampen},  \cite{Arnaudon17}, \cite{Apte2007}, \cite{Cuenod} and \cite{Wilkinson-book}. 
There is extensive research about this topic, in Section \ref{sec:related} we put this work further into context of related research.
The setup \eqref{eq:obs-scheme} includes the practically relevant case that a highly non-linear process $X$ is only observed at start time $0$ and at a final time, say $1$,  see for instance \cite{arnaudon2020diffusion} for an application in shape analysis.

\subsection{Data augmentation for continuous-discrete smoothing }\label{sec:understanding} 
In this section we give intuition on the method that we will give for sampling from the smoothing distribution. 
Assume for the moment that $x_0$ is known and denote $x_{t_i}$ by $x_i$.  Using Bayesian notation (writing $\pi$ for a density), we have the following recursive relation
\begin{align*} \pi(x_{1}, \dots, x_{i} \mid \scr{V}_1) & =
 \pi(x_{i} \mid x_{i-1},  \scr{V}_1) \pi(x_{1}, \dots, x_{i-1} \mid \scr{V}_1).
\end{align*}
Here the first term on the right hand side captures the dynamics of the conditional process, which satisfies
\[  \pi(x_{i} \mid x_{i-1},  \scr{V}_1) = \pi(x_{i} \mid x_{i-1},  \scr{V}_i)  =  \pi(x_i \mid  x_{i-1}) w_i, \]
  where we write suggestively
\[ w_i =   
  \frac{\pi( \scr{V}_i \mid x_{i})}{\pi( \scr{V}_i \mid x_{i-1})}=\exp\left(\log \pi( \scr{V}_i \mid x_{i}) - \log \pi( \scr{V}_i \mid x_{i-1}) 
\right).
\]
In the (non-Bayesian) notation used in the next subsection, $\pi(\scr{V}_i\mid x_i)$, the likelihood of $x_i$ given the  present and future  observations $\scr{V}_i$, is denoted as $\rho(t_i, x_i)$. Up to a scaling factor, $\rho$ can be related to a backward filtered marginal density of $X$.
In a simplified view of our approach, firstly we compute the scaled filtering density backward in time.

One can now show using the arguments laid out in Appendix \ref{sec:doob_h_transf} that the weight equals  the exponential process of a change of measure:
\[
w_i =  \exp\left(\int_{t_{i-1}}^{t_i} \sigma(s,X_s)^\T\nabla \log  \rho(s,  X_s) \dd W_s - \frac12 \int_{t_{i-1}}^{t_i} 
\|\sigma(s,X_s)^\T\nabla \log  \rho(s, X_s)\|^2  \dd s\right),
\]
 which  reveals the changed drift $ b +  \sigma\sigma^\T \nabla \log  \rho$ of the conditional (smoothed) process by Girsanov's theorem.
Thus we may substitute sampling $X_{i} \mid X_{i-1}, \scr{V}_{i}$ by simulating a process  $X$ with drift $b +  \sigma\sigma' \nabla \log  \rho$, understood as \emph{data augmentation} in the sense of \cite{TannerWong}, and obtain a sample of the smoothing distribution of $X_i$ (and in fact the entire smoothed segment $\{X_t, t_{i-1} \le t \le t_i\}$) in this way.

Finally, if $\rho$ cannot be computed efficiently, we find a substitute $\tilde\rho$ and correct for the difference through Monte Carlo methods:  this is the technique of guided proposals as introduced in the following section.

%%%%%%%%%%%%%%%%%%%%%%

\subsection{Approach}\label{sec:approach}
Conceptually, we follow closely the literature on {\it guided proposals}. We explain the main idea here, more technical details can be found in Appendix \ref{sec:recapguided}.

Assume $X$ admits smooth transition densities $p$, i.e.\ for $s<\tau$,  $\mathrm{P}^{(s,x)}(X_\tau \in \!\dd y) = p(s,x; \tau ,y) \dd y$. For $i \in \{1, \dots n\}$,  $t\in (t_{i-1},t_i]$, the process $X$, conditioned on $V_i,\ldots, V_n$ and $X_{t_{i-1}}=x_{t_{i-1}}$  satisfies the SDE 
\begin{equation}\label{eq:form-xstar}
  \dd X^\star_t = b(t,X^\star_t) \dd t+a(t,X^\star_t)r(t,X^\star_t) \dd t+\si(t,X^\star_t) \dd W_t, \qquad X^\star_{t_i}=x_{t_i},
\end{equation}
with   $a(t,x)=\si(t,x)\si(t,x)^\T$,  $r(t,x)=\nabla_x \log \rho(t,x)$ (considered as a column vector) and the likelihood term $\rho$  defined  by integrating out the latent future states,
\begin{equation}\label{eq:generalpull-p}
	 \rho(t,x)=\int p(t,x; t_i, \xi_i)k_n( \xi_n; v_n ) \prod_{j=i}^{n-1} p(t_j, \xi_j; t_{j+1}, \xi_{j+1}) k_j(  \xi_j;  v_j) \dd \xi_i\cdots \dd\xi_n.
\end{equation}
Define $\rho(0,x) = k_0(x; v_0) \rho(0+, x)$. If $X_{t_0} \sim \pi^\star$, with 
\begin{equation}\label{eq:pistar0}
\pi^\star(x_0) = \frac{\pi(x_0) \rho(0, x_0)}{\int \pi(x_0) \rho(0, x_0) \dd x_0}
\end{equation}
being the smoothed (conditional) density of $x_0$, then a trajectory $X^\star$ constitutes a sample from the smoothing distribution. 
 This expression was derived in \cite{Marchand} (Theorem 2.3.4) in a special case of \eqref{eq:lin-obs} with $\Sigma_j\equiv 0$ for all $j$). In Appendix \ref{sec:doob_h_transf} we derive it using Doob's $h$-transform for the more general setting considered in this paper.
As $p$ is tractable only in very specific instances, so are $\rho$ and $\pi^\star$. Hence, sampling $X^\star$ directly from \eqref{eq:form-xstar} is generally not possible. The key idea in \cite{Schauer} (which addressed a simpler problem of diffusion bridge simulation) is to introduce the so-called {\it guided proposal process}, defined as the  (strong) solution to the SDE
\begin{equation}
  \label{eq:Xcirc}
  \dd X^\circ_t = b(t,X^\circ_t) \dd t+a(t,X^\circ_t)\tilde{r}(t,X^\circ_t) \dd t+\si(t,X^\circ_t) \dd W_t, \qquad X^\circ_{t_i}=x_{t_i},
\end{equation}
where $\tilde{r}(t,x)=\nabla_x \log \tilde\rho(t,x)$ and $\tilde\rho(t,x)$ is defined in a similar manner as $\rho(t,x)$, but with  $p$ and $\{k_i\}$ replaced by tractable approximations $\tilde p$ and $\{\tilde k_i\}$ respectively. Here $\tilde p$ can be taken to be the  transition density of some auxiliary diffusion $\tilde{X}$ and $\{\tilde k_i\}$ can be based on \eqref{eq:lin-obs}. The choice of  $\tilde{X}$ and $\{\tilde k_i\}$ impacts the quality with which the law of $X^\circ$ approximates the law of the target process $X^\star$. Intuitively, the drift and dispersion of $\tilde{X}$ should be chosen such that $\tilde{X}$ is similar to $X$ in areas visited by the true conditional process. At the same time, the two functions need to be simple enough, so as to make it possible to compute $\tilde{\rho}(t,x)$ and $\tilde{r}(t,x)$.
  \cite{Schauer} showed that the linear processes, defined by the SDE
\begin{equation}\label{eq:sde-linproc}
  \dd \tilde{X}_t = \left(\beta(t) + B(t) \tilde{X}_t\right) \dd t + \tilde\sigma(t) \dd W_t,
\end{equation}
give rise to a family of auxiliary diffusions that grant a generous degree of flexibility  (as detailed in Section \ref{sec:choice}) for crafting suitable proposals $X^\circ$. 
For instance, in \cite[Section 1.3]{Schauer} the authors showed how in a setting of a highly non-linear target diffusion $X^\star$, a simple shift by a constant in $\beta(t)$ may lead to a dramatic improvement of simulated proposal bridges $X^\circ$, leading e.g.\  to lowering of rejection rates in Metropolis--Hastings algorithms.

Assume for now that  $x_0$ is  known and  $k_i =\tilde k_i$ with $\tilde k_i$ derived from observation scheme \eqref{eq:lin-obs} and transition density $\tilde p$ derived from \eqref{eq:sde-linproc}. We defer the general case to Section \ref{sec:smoothing-algorithm}. Let $\P^\star$ and $\P^\circ$ denote the laws of $X^\star$ and $X^\circ$ on $[0,t_n]$ (started at $x_0$) respectively. Under regularity conditions on the auxiliary process $\tilde{X}$ listed in Appendix \ref{sec:recapguided},  it follows from Theorem 1 in \cite{Schauer} and Theorem 2.14 in \cite{Bierkens} that 
\begin{equation}
\label{eq:goal}	\frac{ \dd\P^\star}{\dd\P^\circ}({X^\circ}) =
 \frac{\tilde \rho(0+,x_0)}{\rho(0+,x_0)} \Psi(X^\circ), 
\end{equation}
where 
\begin{equation}
	\label{eq:Psi}
	\Psi(X^\circ)= \exp\left(\int_0^{t_n} G(s,X^\circ_s) \dd s \right), 
\end{equation}   
and
\begin{equation}\label{eq:defG} \begin{split} G(s,x) &= (b(s,x) - \tilde b(s,x))^\T \tilde r(s,x) -  \frac12 \trace\left(\left[a(s, x) - \tilde a(s)\right] \left[H(s)-\tilde{r}(s,x)\tilde{r}(s,x)^\T\right]\right).
\end{split}
\end{equation}
 The factor in front of $\Psi$ is written to indicate that in this expression $\rho$ and $\tilde\rho$ are to be evaluated at time $0+$, which is of importance when considering the case where  $x_0$ is not known.
Here, $\tilde{b}$ and $\tilde{\sigma}$ are the coefficients of the SDE defining $\tilde{X}$,  $\tilde{a}(s)=\tilde\sigma(s)\tilde\sigma(s)^\T$ and $H(s)$ is the negative of the Hessian matrix of $x\mapsto \nabla_x \log \tilde{\rho}(s,x)$   (which turns out not to depend on $x$).

Since all terms in \eqref{eq:Xcirc} and \eqref{eq:goal} are in principle tractable, one may define an importance sampler that targets the law $\P^\star$: proposals are drawn from $X^\circ$ (by forward simulating paths via some discretisation scheme for SDEs, say, the Euler--Maruyama scheme) and importance weights are computed based on \eqref{eq:goal}. As the SDE for $X^\circ$ is obtained from the SDE for $X^\star$ by superimposing a guiding term, the paths distributed according to the former law are called {\it guided proposals}. The process $X^\circ$ is typically used as a proposal (the origin of the terminology ``guided proposal'') in a ``latent'' path-space Metropolis--Hastings step such as the non-centred path-space Crank-Nicolson scheme (cf.\ \cite{cotter2013} and \cite{beskos-mcmc-methods}).

However, our results can also be used in other Monte Carlo procedures, for example in an auxiliary particle filter. 
	Essentially our backward filtering step is an approximation to the backwards information filter. Various related ``optimal policy finding'' strategies for finding a good approximation have been proposed in the particle filtering literature, such as the iterated auxiliary particle filter (\cite{Guarniero2017})  and controlled SMC (\cite{Heng2020}).
Guided proposals can also be used to create flexible variational classes for latent diffusion paths (cf.\ recent work \cite{beskos2021scorebased}). In fact guided proposals are agnostic to the choice of inferential algorithm, a point highlighted in the preprint \cite{vandermeulen2021automatic}.

Whereas the computations of $\tilde{r}$, $H$ and $\tilde\rho(0,x_0)$ for \eqref{eq:sde-linproc} are in principle tractable, it is not trivial to derive an efficient numerical scheme. The numerical methods presented in \cite{Schauer} are restricted to simpler cases and  to diffusion bridge simulation on a single time segment. {\it In this paper we derive an efficient scalable scheme (both in the number of observations as dimension of the state space of the diffusion) using guided proposals for sampling    high-dimensional, non-linear latent diffusion processes under the general observation scheme \eqref{eq:obs-scheme}, and to estimate unknown parameters.} This is illustrated in a number of challenging numerical examples. 

\subsection{Innovation} 
We derive simple systems of ordinary differential equations that can be used to compute all terms needed for implementation of guided proposals, i.e. terms $\tilde{r}$, $H$ and $\tilde\rho(0,x_0)$.
These results imply that in order to sample paths of the proposal diffusion $X^\circ$ and embed them in the Metropolis--Hastings algorithm only the following steps need to be performed
 \begin{itemize}
  \item a  system of ordinary differential equations, akin to the updating equations in hybrid Kalman filtering, has to be solved backwards in time (suitable systems are given in theorems \ref{thm:rhotilde_using_MLmu}, \ref{thm:FHc_ODEs} and \ref{thm:noise-Hd});
  \item paths of $X^\circ$ need to be simulated forward in time, using standard simulation techniques based on stochastic Taylor expansions (cf.\ \cite{KloedenPlaten});
  \item the Radon--Nikodym derivative between the laws of $X^\star$ and  $X^\circ$ has to be evaluated and then used in the acceptance probability of the Metropolis--Hastings algorithm.
\end{itemize}
In particular, in this scheme, forward simulating the guided proposal $X^\circ$ is comparable in computational effort to simulating $X$ itself. This resolves objections about (perceived) computational effort required to simulate guided proposals using a closed form expression for $\rho$. The systems of ODEs  make use only of the matrix addition, multiplication and inversion operations and scale much better to high dimensional problems compared to for example \cite{vdm-schauer}. For the smoothing problem, the first step needs to be executed only once.

We call the resulting algorithm the {\it Backward Filtering Forward Guiding} (BFFG) algorithm, because (as we will show) it combines backward filtering under the auxiliary process $\tilde{X}$ and forward simulation of the guiding process $X^\circ$.  
It may also be characterised as a variant of the forward filtering-backward sampling algorithm for linear state space models with (possibly) non-linear continuous time processes. Reversing the time-direction and sampling the process forward in time is not strictly necessary, but turns out to be more convenient.

We believe the method derived in this paper has a number of attractive properties:
\begin{enumerate}
\item It is a computationally simple algorithm that provides a unified approach to smoothing of both hypo-elliptic  and uniformly elliptic diffusions.
\item It allows for taking into account nonlinearities in the drift efficiently  (by choice of the auxiliary process $\tilde{X}$).
\item It can deal with a large class of diffusions with state-dependent diffusion coefficient.\item The algorithm targets the exact smoothing distribution and does not rely on Gaussian approximations. 
\item I can deal with nonlinear observation schemes or non-Gaussian error densities while still sampling from the exact smoothing distribution.
\end{enumerate}
Regarding being exact, we derive our algorithm in continuous time, but ultimately, in any implementation the SDE for $X^\circ$ needs to be discretised. However, the mesh-width for the discretisation can be controlled by the user.
 
\subsection{Outline}

In Section \ref{sec:backwardfiltering} we derive ODE-systems for backward filtering which result in efficient ways for computing $\tilde r$, $H$ and $\tilde\rho(0,x_0)$.  In  Section \ref{sec:auxiliary}  we discuss strategies for choosing  the auxiliary process $\tilde X$ and auxiliary observation scheme to be used in backward filtering.  The backward filtering forward guiding algorithm for diffusions is subsequently given in Section \ref{sec:smoothing-algorithm}.  In Section \ref{sec:high} we address applications of the proposed algorithms to problems in which the dimension of the state-space of the diffusion is high. In such a setting, the backward filtering can for example be carried out using an ensemble Kalman filter. 
We illustrate our results on 3 challenging examples in Section \ref{sec:examples}. The appendix contains sections on related literature, a derivation of the guiding term for guided proposals together with deferred proofs and remarks.

%--------

%%%%%%%%%%%%%%%%%%%%%%

%------
\section{Backward filtering}\label{sec:backwardfiltering}

To  sample the guided proposal, we need to define  $\tilde\rho$ as  a proxy to $\rho$ defined in \eqref{eq:generalpull-p}. For that, we choose $\tilde X$ as in \eqref{eq:sde-linproc}.
For $i=0,\ldots, n$, we assume $\tilde k_i$ to correspond to the observation scheme \eqref{eq:lin-obs} which is parametrised by $L_i$ and $\Sigma_i$. 
\begin{defn}
 For a specified (strictly positive definite) covariance matrix $P_{n+}$, define $\tilde k_{n+}(x) = \phi(x; 0, P_{n+})$.  We define $\tilde\rho$  by 
\begin{equation}\label{eq:generalpull-ptilde}
	 \tilde\rho(t,x)=\int \tilde k_{n+}(\xi_n)	 \tilde p(t,x; t_i, \xi_i) 	 \tilde k_n( \xi_n; v_n ) \prod_{j=i}^{n-1} 	 \tilde p(t_j, \xi_j; t_{j+1}, \xi_{j+1}) 	 \tilde k_j(  \xi_j;  v_j) \dd \xi_i\cdots \dd\xi_n.
\end{equation}
\end{defn}
Note the inclusion of $k_{n+}$ which does not appear in \eqref{eq:generalpull-p}. It ensures regularisation, where we may intuitively think of $P_{n+} = \eps^{-1} I$ with $\epsilon$ small, though any choice of $P_{n+}$ will give rise to a valid algorithm targeting the exact smoothing distribution.

Naturally BFFG will work best if  $\{\tilde k_i\}$ and $\tilde X$ approximate $\{k_i\}$ and $X$ well, most importantly in those areas where the smoothing distribution puts its weight. 
We comment on  good choices in Section \ref{sec:auxiliary}. 
 Theorems \ref{thm:rhotilde_using_MLmu}, \ref{thm:FHc_ODEs} and \ref{thm:noise-Hd} in this section  give backward recursions for evaluating $\tilde r$, $H$ and $\tilde\rho$ which are needed for application of guided proposals. All proofs are in Section \ref{sec:proofs} of the appendix.

We impose 
 \begin{ass}\label{ass:SigmaT}
For each $i\in \{0,\ldots,n\}$, the matrix $\Sigma_i$ is strictly positive definite. 
 \end{ass}
\begin{ass}\label{ass:existence}
	The maps $t\mapsto \beta(t)$, $t\mapsto B(t)$ and $t\mapsto \tilde\sigma(t)$ are bounded on $[0,T]$. 
\end{ass}
By Theorem 1.184 in \cite{chicone2006ordinary}, Assumption \ref{ass:existence}  ensures existence and uniqueness of the ODEs for $t\mapsto L(t)$, $t\mapsto P(t)$ and $t\mapsto \nu(t)$ appearing in Theorem \ref{thm:rhotilde_using_MLmu} and Theorem \ref{thm:noise-Hd} below.
Uniqueness and existence of $\nu$ and $P$ then translates to the ODEs for $t\mapsto F(t)$ and $t\mapsto H(t)$ appearing in Theorem \ref{thm:FHc_ODEs} below. 

For $i\in \{1,\ldots, n\}$, let $v(t)$ be defined by the concatenated vector of all non-past observations:
\[ v(t) = [ v_i^\T, \ldots, v_n^\T ]^\T  \qquad t\in (t_{i-1},t_{i}], 
\]
where $i\in \{1,\ldots, n\}$. Define $v(0)=[ v_0^\T, \ldots, v_n^\T ]^\T$,  
let $m(t) = \mbox{dim}(v(t))$ and recall $m_i = \mbox{dim}(v_i)$. 
In the following, we give three results characterising
 $\tilde\rho$, first directly as marginal likelihood, and then in a form more suitable for implementation: as (scaled) backward ODE for the filtering density in two parametrisations.
\begin{thm}\label{thm:rhotilde_using_MLmu}
Let the triplet $(L(t), \Md(t), \mu(t))$ be defined on each interval $(t_{i-1},t_i]$ ($i=1,\ldots, n$) as solutions to the backward ODE systems
\begin{align}
 \dd L(t) &= -L(t) B(t)\dd t \label{eq:ode-tildeL} \\
 \dd \Md(t)&=- L(t) \tilde{a}(t) L(t)^\T\dd t \label{eq:ode-Mdagger} \\
 \dd \mu(t) &=-L(t) \beta(t)\dd t. \label{eq:ode-mu}
\end{align}
Define
\[ L(t_n+)=I \quad \Md(t_n+)=P_{n+} \quad \mu(t_n+)=0 \]
and for $i=0,\ldots, n$, 
\begin{equation}\label{eq:change-at-S} L(t_i)=\Bm L_i \\ L(t_i+) \Em, \qquad 
 \Md(t_i)=\Bm \Sigma_i & 0_{m_i\times m(t_i+)} \\ 0_{m(t_i+)\times m_i} & \Md(t_i+)\Em,  \qquad 
 \mu(t_i)=\Bm 0_{m_i\times 1} \\ \mu(t_i+) \Em. 
\end{equation}
If $M(t)=\Md(t)^{-1}$, then for all $t\in [0,t_n]$, we have 
\begin{align*}
 \tilde{r}(t,x) &= F(t) - H(t) x,	\\
  \tilde{\rho}(t,x)& = \phi(v; \mu(t)+L(t) x, \Md(t)) 
\end{align*} 
where
\begin{align} H(t)&=L(t)^\T M(t) L(t) \label{eq:tildeH}, \\
F(t) &= L(t)^\T M(t) (v(t)-\mu(t))\label{eq:F}.
\end{align}
\end{thm}
%%%%%%%%%%

Note that the equations for $\Md$ and $\mu$ directly give the time derivative and that the values of $\Md(t)$ and $\mu(t)$ can be found by using a numerical quadrature rule such as for example the trapezoid rule. This theorem shows that $H$ and $\tilde{r}$ can be computed by solving the systems \eqref{eq:ode-tildeL}, \eqref{eq:ode-Mdagger} and \eqref{eq:ode-mu}. Additionally, by computing the term $|M(t)|$, we may evaluate $\tilde\rho$ as well. In Theorem \ref{thm:FHc_ODEs} below we will see that computation of this determinant can be avoided by solving another (scalar-valued) backward ODE. 

The  dimensions of the  equations  in Theorem \ref{thm:rhotilde_using_MLmu} are larger on $(t_{i-1},t_i]$ than on $(t_{j-1},t_j]$ for all $j>i$. The dimension on the segment closest to zero (i.e.\ the leftmost segment) grows rapidly with a large number of observations.
 It turns out that other sets of backward differential equations  can be derived, which are constant in dimensions determined by the the dimension of the state space of the diffusion, no matter the number of observations. 
 
 \begin{thm}[Information filter]\label{thm:FHc_ODEs}
For $(t,x) \in [0,T]\times \RR^d$,  $\tilde{\rho}$ admits the following decomposition
    \begin{equation*}
\log \tilde{\rho}(t,x) = - c(t) -\frac{1}{2} x^\T  H(t) x +F(t)^\T x    ,
    \end{equation*}
where   on each interval $(t_{i-1},t_i]$ ($i=1,\ldots, n$),   $H$, $F$ and $c$ solve the backward ODEs 
\begin{equation}
  \label{eq:backwardODE}
    \begin{split}
        \dd H(t) &= \left( -B(t)^\T H(t) - H(t)B(t)+H(t)\tilde{a}(t)H(t) \right)\dd t,\\
        \dd F(t) &= \left( -B(t)^\T F(t) + H(t)\tilde{a}(t)F(t) + H(t)\beta(t) \right)\dd t,\\
        \dd c(t) &= \left( \beta(t)^\T F(t) + \frac{1}{2}F(t)^\T \tilde{a}(t)F(t) - \frac{1}{2}\trace\left(H(t)\tilde{a}(t)\right) \right)\dd t.
    \end{split}
  \end{equation}
At observations times, we have for  $i=0,\ldots, n$
\begin{equation}
  \label{eq:updateODE}
    \begin{split}
    H(t_i) &= H({t_i+})+ L_i^\T \Sigma_i^{-1}L_i,\\
    F(t_i) &= F({t_i+})+L_i^\T \Sigma_i^{-1}v_i,\\
    c(t_i) &= c({t_i+}) - \log \phi(v_i; 0,\Sigma_i).
    \end{split},
  \end{equation}
 to be initialised from 
 \begin{equation}\label{eq:initHFc} H(t_n+) = P_{n+}^{-1}, \quad  F(t_n+) =0, \quad   c(t_n+) = \phi(0; 0, P_{n+}) \end{equation} 
\end{thm}
Note that $H(t) \in \RR^{d\times d}$, $F(t) \in \RR^d$ and that $c(t)$ is scalar-valued throughout $[0,t_n]$.

\begin{thm}[Covariance filter]\label{thm:noise-Hd}
Let $\nu(t)= P(t) F(t)$ with $\Hd(t)=H(t)^{-1}$. 
The mapping $\tilde{r}$ satisfies
\begin{equation}\label{eq:def_rtilde_in_H_and_nu}
P(t) \tilde{r}(t,x) =  \nu(t)-x,
\end{equation}
On each interval $(t_{i-1},t_i]$ ($i=1,\ldots, n$), $\Hd$ and $\nu$ satisfy the backward ODEs
\begin{align}
\dd \Hd(t) &= \left( B(t) \Hd(t) + \Hd(t) B(t)^\T - \tilde{a}(t) \right) \dd t, \label{eq:recP}  \\
\dd \nu(t) &= \left( B(t) \nu(t) + \beta(t)\right) \dd t.\nonumber
\end{align}
Furthermore, if we define $P(t_n+) = P_{n+}$ and $\nu(t_n+)=0$ then 
for $i=0,\ldots, n$ 
\begin{align}  \Hd(t_i) &= \Hd(t_i+) - \Hd(t_i+)L_i^\T \left(\Sigma_i + L_i \Hd(t_i+) L_i^\T\right)^{-1} L_i \Hd(t_i+)   \label{eq:rec-Hd-init}  \\
 \nu(t_i) &=\Hd(t_i) \left(L_i^\T \Sigma_i^{-1} v_i +H(t_i+)\nu(t_i+)\right).\label{eq:rec-nu-init}\end{align}
\end{thm}
From \eqref{eq:rec-Hd-init} it is seen that Assumption \ref{ass:existence} can be weakened to invertibility of $\Sigma_i + L_i P(t_i+)L_i^\T$. 

In case $\tilde{\sigma}$ and $B$ are constant, $P(t)$ can be computed directly. 
\begin{prop}\label{prop:lyap} Assume $B$ and $\tilde{a}$ are constant  and $\Sigma$ solves the continuous time Lyapunov equation 
\[B \Sigma +  \Sigma B^\T + \tilde{a} = 0.\]
Then the solution to 
\[
\dd \Hd(t) = \left( B \Hd(t) + \Hd(t) B^\T - \tilde{a} \right) \dd t,\qquad \Hd(T)=\Hd_T
\]
is given by
\begin{equation}\label{eq:Pt}
\Hd(t) =  
   \Phi(t, T)(\Sigma+\Hd_T)\Phi(t, T)^\T -  \Sigma, 
\end{equation}
where $\Phi(t, T) = \exp(-(T-t)B)$.
\end{prop}

\begin{rem}
It is natural to ask which of the 3 recursions to use. In case of only one future observation at time $t_1$, the ODEs in Theorem \ref{thm:rhotilde_using_MLmu} are advantageous when $m_1<d$. Moreover, these are also applicable in case of a noiseless observation at time $t_1$ by taking $\Md(t_1)$ to be equal to a zero matrix. 
With multiple observations, usually, the recursions from Theorem \ref{thm:FHc_ODEs} or Theorem \ref{thm:noise-Hd} are preferable and can lead to large computational gains. 
\end{rem}

%%%%%%%%%%%%%%%%%%%%%%

%\subsection{Interpretation of backward filtering}\label{subsec:filtering}
\begin{rem}
Let $\scr{V}_t = \{V_{i},\ldots, V_n\}$ if $t\in (t_{i-1}, t_i]$. 
The interpretation of $\nu$ and $\Hd$ (appearing in Theorem \ref{thm:noise-Hd}) follows from $\tilde{X}_t \mid \scr{V}_t \sim \operatorname{N}(\nu(t), P(t))$, i.e.\ these are the mean and covariance of the backward filtered process. The update formula at observations times can now be seen from the following argument. Denote $\tilde{X}_i \equiv \tilde{X}_{t_i}$. 
  Suppose $(\nu(t), P(t))$ has been computed for $t\in (t_i, t_n]$ for $i\in \{0,\ldots, n-1\}$. 
At time $t_i$ the observation $\tilde{X}_i$ can be incorporated using the following relation:
\begin{equation}\label{eq:update_obs} p(\tilde{X}_{i} \mid \scr{V}_i) \propto p(V_i \mid \tilde{X}_i, \scr{V}_{i+1}) p(\tilde{X}_i \mid \scr{V}_{i+1})=p(V_i \mid \tilde{X}_i) p(\tilde{X}_i \mid \scr{V}_{i+1}) .
\end{equation}
Since $\tilde{X}_i \mid \scr{V}_{i+1}\sim \N(\nu(t_i+), P(t_i+))$  and $V_i \mid \tilde{X}_i \sim \operatorname{N}(L_i \tilde{X}_i, \Sigma_i)$, the joint distribution of $(\tilde{X}_i, V_i) \mid \scr{V}_{i+1}$ is   Normal. Hence there is a closed form expression for the distribution of $\tilde{X}_i \mid (V_i, \scr{V}_{i+1})$ which gives $\nu(t_i)$ and $\Hd(t_i)$. 
 
 \end{rem}

\section{Choice in auxiliary process and auxiliary observation scheme}\label{sec:auxiliary}
Application of the tractable backward filtering updates requires choosing $(\{L_i\}, \{\Sigma_i\})$ for the observation scheme and $t\mapsto(\beta(t),  B(t), \tilde\sigma(t))$ for the auxiliary process $\tilde X$.

\subsection{Choice of the auxiliary observation scheme}\label{subsec:bfnonlinear}

The backward filtering formulas in Section \ref{sec:backwardfiltering} are used with the auxiliary observation scheme of the form implied by \eqref{eq:lin-obs}, leaving open the choice of $(L_i, \Sigma_i)$ in case the actual observation scheme is of the more general form \eqref{eq:obs-scheme}. 

The simplest thing to do is choose $L_i$ and $\Sigma_i$ such that 
$\phi(v_i; L_i X_{t_i}, \Sigma_i) \approx k_i(X_{t_i}; v_i)$. Ultimately, the best choice of  $(L_i, \Sigma_i)$ is  that the discrepancy between $\P^\circ$ and $\P^\star$ is minimal (for example measured by (reverse) Kullback-Leibler divergence).
\begin{ex}
Assume that instead of $V_i \sim \N(L_i X_{t_i}, \Sigma_i)$ one observes $V_i \sim \N(g_i(X_{t_i}), \Sigma_i)$ with $g_i$ a nonlinear map. Consider the linearisation (with $\nu$ as in the covariance filter)
\[ g_i(x) \approx g_i(\nu(t_i+)) + G_i (x-\nu(t_i+))=\zeta_i+ G_i x. \]
where $G_i = (Dg_i)(\nu(t_i+))$ and $\zeta_i = g_i(\nu(t_i+))-G_i \nu(t_i+)$. Under this linearisation, backward filtering can be carried out exactly as in our original setting by assuming to observe $V_i-\zeta_i$ instead of $V_i$ and setting $L_i$ to be equal to $G_i$. 
This choice of auxiliary scheme is alike the extended Kalman filter	 (Algorithm 5.4 in \cite{Sarkka-book}).
\end{ex}

 We refer to Chapters 5 and 6 in \cite{Sarkka-book} for other approximations proposed in the literature that can be used as auxiliary observation scheme. 

\subsection{Choice of the auxiliary process $\tilde{X}$}\label{sec:choice}
The auxiliary process $\tilde{X}$ needs to be chosen by a user, i.e.\ the parameters $\beta(t)$, $B(t)$ and $\tilde\sigma(t)$ need to satisfy matching and regularity conditions, but are otherwise free. A number of practical ways for choosing them are discussed in \cite{vdm-s-estpaper} (Section 4.4); below, we list a number of reasonable options. Determining an optimal auxiliary law in an automated fashion deserves its own study and is the subject of an ongoing research (but cf. also considerations in \cite{Schauer}).

\begin{enumerate}[label=\textbf{\Alph*}]
\item\label{enum:aux_zero_drift} Waive the freedom and simply take  $\tilde\sigma$ constant, $B(t)\equiv 0$  and $\beta = 0$.  If $X$ is either elliptic or hypo-elliptic with nonzero noise on each coordinate, this yields a valid algorithm. It still takes local nonlinearity into account through the presence of $b$ in \eqref{eq:Xcirc} and, in case of partial observations, is informed by multiple future observations. 
\item\label{enum:aux_linear_at_end} On each $t\in (t_{i-1},t_i]$ compute a first order Taylor expansion of the drift evaluated at a specified $\tilde{x}(t_i)$. That is, compute $\tilde b(t, x) = b(t, \tilde{x}(t_i)) + J_b(t,\tilde{x}(t_i)) (x - \tilde{x}(t_i))$, where $J_b$ denotes the Jacobian matrix of $b$. In case the diffusion is fully observed,    $\tilde{x}(t_i)$ is taken equal to $X_{t_i}$, else it is specified by a user-chosen guess for it. The guess can be adaptively refined by using empirical averages of the imputed path. Otherwise put, the auxiliary process comes from linearisations of the target diffusion at the times of subsequent observations.
\item\label{enum:aux_linear_at_mean} The aim is to employ a first order Taylor expansion $\tilde b(t, x) = b(t, \bar x(t)) + J_b(t,{\bar x(t)}) (x - \bar x(t)) $ evaluated at ${\bar x(t)}= \expec{X_t \mid D}$. Of course, $\bar{x}(t)$ is unknown, but information becomes available during MCMC-iterations, and thus, just as in \ref{enum:aux_linear_at_end}, empirical averages of the imputed path can be used as its proxy.
More specifically, we propose to first use  the auxiliary process constructed by one of the preceding methods to get $B^{(0)}(t,x)=\beta(t)+B(t) x$, and then, run the algorithm for $k$ iterations and compute the average of these $k$ paths.  Denote this average by $\{\bar{X}^{(0)}(t),\, t\in [0,t_n]\}$. Next, starting at $i=1$, repeat the following steps until the prescribed total number of iterations for adaptation has been reached.
\begin{enumerate}
  \item  Set
\[ B^{(i)}(t,x) = b(t,\bar{X}^{(i-1)}(t)) + J_b(t, \bar{X}^{(i-1)}(t)) \left(x-\bar{X}^{(i-1)}(t)\right). \]

\item Recompute $H(t)$, $F(t)$ and $c(t)$ on $[0,t_n]$ based on $B^{(i)}(t,x)$. 
\item Perform $k$ MCMC-iterations based on $H(t)$, $F(t)$ and $c(t)$. Compute the average of all simulated paths to obtain 
$\{\bar{X}^{(i)}(t),\, t\in [0,t_n]\}$
\end{enumerate}

To avoid common problems with adaptive schemes we stop the adaptation after a fixed number of steps.

\item\label{enum:aux_const_drift} Choose $\tilde\sigma=\sigma$, $B(t)\equiv 0$  and take  $\beta$ nonzero to make the pulling term itself take into account the nonlinearity of the system.  To determine $\beta$, first obtain $F(t_n+)$ and $H(t_n+)$ as in \eqref{eq:initHFc}.
Next, for $i=n$ to $1$:
\begin{enumerate}
  \item For $t\in (t_{i-1},t_i]$, backwards solve 
\[ \dd x(t) =b(t,x(t)) \dd t,\qquad x(t_i)=H(t_i)F(t_i). \]
Set $\beta(t)=b(t,x(t))$.
\item Using $\beta$, compute $(F(t), H(t))$ for all  $t\in (t_{i-1},t_i]$. Compute $(F(t_{i-1}), H(t_{i-1}))$ using $(F(t_{i-1}+), H(t_{i-1}+))$  using the formulas of the information filter (Theorem \ref{thm:FHc_ODEs}). 
\end{enumerate}

\item\label{enum:aux_linear_comb} Use a linear combination of the schemes above. This is a risk-averse, robust strategy, which aims to guard against unexpected artifacts that some of the adaptive strategies above might exhibit (say, for strategy \ref{enum:aux_linear_at_mean} such behaviour could appear as a result of multimodality on a path space). In Section \ref{sec:examples}  we use: $B=wb^{(B)}+(1-w)b^{(C)}$, where $b^{(B)}$ and $b^{(C)}$ are defined by strategies \ref{enum:aux_linear_at_end} and \ref{enum:aux_linear_at_mean} respectively and $w\in[0,1]$ are some user-chosen weights. Another option (not considered here) consists of sampling $w$ in each iteration from the Bernoulli distribution with fixed success probability.

\item Finally, the auxiliary process may take values in a higher dimensional space, with only its first $d$ coordinates used in defining the proposal diffusion. Say, we want to target a process $X$ conditioned on the observations $V_i = L X_{t_i} + \eta_i$. Consider an auxiliary process $[\tilde X,Y]$ with $\tilde X$ solving $\dd \tilde X_t = (B \tilde X_t + \beta(t) + C Y_t) \dd t + \tilde \sigma \dd W_t$ and $Y$ a second, independent linear process $\dd Y_t = D Y_t\dd t + \dd W^Y_t$ 
  driven by an independent Brownian motion $W^Y$.  Here $B, C$ and $D$ are matrices of suitable dimension. Then, we may use an augmented sampler that targets the joint process $[X_t, Y_t]$ 
conditional on the observations $V_i$. The augmented observation operator is given by $L_{aug} = [L\; 0]$, the augmented drift for the target process is $b_{aug}(t, [x,y]) = [b(t,x), Dy]$ and the block diagonal dispersion coefficient is $\left[\begin{smallmatrix} \sigma & 0\\ 0 &I\end{smallmatrix}\right]$; the sampler uses the augmented guided proposal with the auxiliary linear process $[\tilde X, Y]$ that has a drift $\tilde b_{aug}(t, [x, y]) = [B x + C y, D y]$ and the block diagonal dispersion coefficient $\left[\begin{smallmatrix} \sigma & 0\\ 0 &I\end{smallmatrix}\right]$. Samples of the conditional process $X$ are obtained by marginalisation. To give a simple example, $\tilde X$ can be taken to be a sum of a Brownian motion and an integrated Brownian motion.
\end{enumerate}

%%%%%%%%%%%%%%%%%%%%%%%
%\section{A novel smoothing and parameter estimation algorithm}
\section{Backward Filtering Forward Guiding for diffusions}
\label{sec:smoothing-algorithm}

In this section we present a novel Gibbs sampling algorithm for joint continuous-discrete smoothing of diffusions and parameter estimation.  We will first precisely derive the posterior distribution that we target in Section \ref{subsec:lik}. Next, in Section \ref{subsec:init} we present details on initialisation of the algorithms, whereas steps for updating the path, initial state and parameters and given in sections \ref{subsec:smooth}, \ref{subsec:initial-state} and  \ref{subsec:parameter} respectively. 

\subsection{Likelihood calculations}\label{subsec:lik}

The law $\P^\circ$ of the guided proposal  approximates $\P^\star$ with tractable likelihood ratio.
In the simple setting of known starting point $x_0$ and  $k_i =\tilde k_i$ with $\tilde k_i$ derived from observation scheme \eqref{eq:lin-obs} we have  specified this ratio  in \eqref{eq:goal}. Dropping the requirement that $k_i = \tilde k_i$, we have  for bounded measurable test-functions $g$ 
\begin{equation}\label{eq:gXstar_x0known}
 \EE \left[g(X^\star) \mid X^\star_0=x_0\right] = \frac{\tilde\rho(0+,x_0)}{\rho(0+,x_0)} \EE \left[  g(X^\circ) \Psi(X^\circ) \left(\prod_{i=1}^{n} C_i(X^\circ_{t_i}) \right) \big| X^\circ_0=x_0\right],
\end{equation}
where $C_i(x) = k_i(x; v_i)/\tilde k_i(x; v_i)$ for $1\le i \le n-1$ and $C_n(x) =  k_n(x; v_n)/(\tilde k_n(x; v_n) \tilde k_{n+}(x))$. This follows from \eqref{eq:goal} upon correcting for the discrepancy between $\tilde k_i$ and $k_i$. 

Now we assume $b$, $\sigma$, $\tilde{p}$, $\{\tilde k_i\}$ , $\{k_i\}$, $\pi(x_0)$ depend on an unknown parameter $\theta$ with prior $\kappa(\theta)$. We add a subscript $\theta$ if we wish to highlight this dependency. 

A naive approach for sampling jointly the latent path $X^\star$ and parameter $\theta$ would consist of  data-augmentation where one iteratively updates the latent path $X^\star$ proposing from $X^\circ$ conditionally on the parameter $\theta$ and $\theta$ conditionally on $X^\star$. 
It is well known that this yields a reducible algorithm if unknown parameters appear in the  diffusivity $\sigma$ (cf.\ \cite{RobertsStramer}). The key idea put forward in \cite{MR2422763} (see also \cite{vdm-s-estpaper} and \cite{PapaRobertsStramer}) is to use a noncentred parametrisation where the law of $X^\circ$ is casted as a pushforward of Wiener measure. It is this approach that we adopt  here as well: 
as we assume existence of a strong solution to the SDE \eqref{eq:Xcirc}, there exists of a measurable map $\GP_\th$ such that \[ X=\GP_\th(X_0, Z), \] where $Z$ is a Wiener process in $\RR^{d'}$. 	The process $Z$ is referred to as the {\it innovation process}. 

Below, we define algorithms \ref{mainalg},  \ref{mainalg-initpoint} and \ref{mainalg-parest}. These are to be combined in a Gibbs sampler to target the joint posterior distribution of $(\th, x_0, Z)$ specified by the proper density
\begin{equation}\label{eq:target}
\frac{ \kappa(\theta)  
 \pi_\theta(x_0) \tilde\rho_\th(0, x_0)}{\int \kappa(\theta) \pi_\th(x_0)  \rho_\th(0, x_0)   \dd (x_0 , \theta) }
\Psi_\th(\GP_\th(x_0, Z)) \prod_{i=0}^{n} C_i(\GP_\th(x_0, Z)_{t_i}) ,
\end{equation}
with respect to the product measure of Lebesgue measure on $(\th, x_0)$ and $d'$-dimensional Wiener measure on $Z$. 
In particular, samples of the latent path are obtained from samples $(\theta, x_0, Z)$ through $X=\GP_\theta(x_0, Z)$.

To see why \eqref{eq:target} is true, if $g$ is a bounded measurable function, then, with $\mathbb{W}$ denoting Wiener measure and $X_0$, $\Theta$ random variables drawn from the prior,
\begin{align*}
\EE & [g(\Theta, X_0, Z) \mid \scr{V}_0]  =  \EE [ \EE [g(\Theta, X_0, Z) \mid \scr{V}_0, \Theta, X_0 ]  \mid \scr{V}_0]\\ 
& =  \int \EE [g(\Theta, X_0, Z) \mid \scr{V}_0, X_0= x_0, \Theta=\theta] \xi(x_0, \theta; \scr{V}_0) \dd (\theta, x_0)\\ 
&=
\int  g(\theta, x_0, z)   \xi(x_0, \th; \scr{V}_0)  \frac{\tilde\rho_\theta(0+,x_0)}{\rho_\theta(0+,x_0)} \Psi_\theta(GP_\theta(x_0, z)) \prod_{i=1}^n C_i(\GP_\theta(x_0, z)_{t_i}) \dd \mathbb{W}(z) \dd \theta\, \dd x_0
\end{align*}
where 
\[ \xi(x_0, \theta; \scr{V}_0)  = \frac{\kappa(\theta) \pi_\theta(x_0) k_0(x_0, v_0) \rho_\theta(0+,x_0)}{\int \kappa(\theta) \pi_\theta(x_0) k_0(x_0, v_0) \rho_\theta(0+,x_0) \dd (\theta, x_0)}
\]
is the density of $(x_0, \theta)$ conditional on $\scr{V}_0$. 
Finally, use $\rho_\theta(0,x_0) =  k_0(x_0, v_0) \rho_\theta(0+,x_0)$ and likewise for $\tilde k_0$, $\tilde \rho_\theta$ and multiply both the numerator and denominator by $\tilde k_0(x_0; v_0)$ and ``absorb'' the ratio $k_0(x_0; v_0)/\tilde k_0(x_0;v_0)$ into the product by starting at $i=0$ rather than $i=1$. 

\subsection{Smoothing from known initial state and parameter}\label{subsec:smooth}
 In this conditional update step we assume that the initial state and parameter are either known or conditioned upon. This means sampling from the smoothing distribution which requires  the likelihood ratio of $\P^\star$ with respect to $\P^\circ$ as specified in \eqref{eq:gXstar_x0known}. 
Recall the definition of $\Psi$ in \eqref{eq:Psi}.

For this step, we choose  a tuning parameter $\lambda \in [0,1)$. 

\begin{alg}\label{mainalg} {\it Smoothing step for fixed $\th$ and fixed initial state $x_0$}. Suppose the current iterate is $(\th, x_0, Z)$ and $X=\GP_\theta(x_0, Z)$.  \begin{enumerate}
  \item {\bf Compute the guiding term}. Initialise $H(t_n+)$, $F(t_n+)$, $c(t_n+)$ via \eqref{eq:initHFc}. 
    
 For $i=n$ to $0$
\begin{enumerate}
	\item For $t\in  (t_{i},t_{i+1}]$, backwards solve the ordinary differential equations \eqref{eq:backwardODE}
  \item Compute $H(t_i)$, $F(t_i)$, $c(t_i)$ from \eqref{eq:updateODE}
 \end{enumerate}

\item {\bf Crank-Nicolson step}.
\begin{enumerate} 
 \item  Sample independently a Wiener  process $W$ and set \[Z^\circ=\lambda Z + \sqrt{1-\lambda^2} W.\] Compute  \[X^\circ=\GP_\theta(x_0, Z^\circ).\]
\item Compute 
\[ A = \frac{\Psi(X^\circ)}{\Psi(X)} \prod_{i=1}^{n} \frac{C_i(X^\circ_{t_i})}{C_i(X_{t_i})}. \] Draw $U\sim \scr{U}(0,1)$. If $U<A$  replace  $X=X^\circ$ and $Z=Z^\circ$.
\end{enumerate}
\end{enumerate}
\end{alg}
Repeating step 2 multiple times constitutes a procedure to sample from the smoothing distribution, in which case step 1 only needs to be done once. 
We have used the information filter from Theorem \ref{thm:FHc_ODEs}. Clearly, with straightforward modifications it can be adjusted to other filtering equations from Section \ref{sec:backwardfiltering}.

ODEs for $H(t)$ and $F(t)$ are calculated on $t\in [t_0,t_n]$ using time discretisation. Once these have been calculated and stored (on a grid of timepoints), the remainder of the algorithm consists of preconditioned Crank--Nicolson steps on the Wiener increments. 
After ``burn in'', the sample paths generated by this algorithm are from the smoothing distribution. Note that since we assume the initial state $x_0$ to be known and the parameter $\theta$ to be fixed, $c(t)$, in fact, does not need to be computed. As we will shortly extend the presented algorithm to include uncertatinty over the initial state and parameter estimation we have already included the ODE and the update formula for $c(t)$. 
 
While Algorithm \ref{mainalg}  is theoretically valid for any fixed value of $\lambda$, its efficiency strongly depends on the particular choice of this parameter. Instead of a fixed value of $\lambda$, we can choose it randomly at each iteration of step (3a): the acceptance probability in step (3b) is not affected by its value (cf.\ the discussion after Algorithm 1 in \cite{vdm-s-estpaper}). 

 \begin{rem}
If the measurement error is close to zero, then the (Euler) discretisation of the guided proposal is a delicate matter due to the behaviour of the guiding term just prior to observation times. As discussed in Section 5 of \cite{vdm-s-estpaper} a time change and scaling of the process can alleviate discretisation errors. For completeness, we present this approach in Appendix \ref{sec:timechange}. 
\end{rem}

\subsection{Initial state updating}\label{subsec:initial-state}
 
We choose to update the initial state  conditional on $(Z,\theta)$.

The expression for $A$ follows from the target density in \eqref{eq:target}. We choose a Markov kernel $q$ (which may depend on $\th$) for proposing a new state $x_0^\circ$.
\begin{alg}\label{mainalg-initpoint} {\it Initial state updating step}. Suppose the current iterate is $(\th, x_0, Z)$ and $X=\GP_\theta(x_0, Z)$.
\begin{enumerate}
  \item Propose $x_0^\circ \sim q_\theta(\cdot \mid x_0)$.
	\item Compute the corresponding guided proposal $X^\circ = \GP_\th(x_0^\circ,Z)$.
	\item Compute 
\[ A = \frac{q_\th(x_0 \mid x^\circ_0)}{q_\th(x^\circ_0 \mid x_0)} \frac{\pi(x^\circ_0)\tilde\rho(0,x^\circ_0)}{\pi(x_0)\tilde\rho(0,x_0)} \frac{\Psi(X^\circ)}{\Psi(X)} \prod_{i=0}^{n} \frac{C_i(X^\circ_{t_i})}{C_i(X_{t_i})}. 
\]
 Draw $U\sim \scr{U}(0,1)$. If $U<A$  replace  $X  = X^\circ$ and $x_0=x_0^\circ$.
\end{enumerate}
\end{alg}

An independence sampler can be obtained by taking $q_\theta(\cdot \mid x_0) = \pi^\circ(\cdot)$ defined by
\begin{equation}\label{eq:picirc}	\pi^\circ(x_0) = \frac{\tilde\pi(x_0) \tilde\rho(0, x_0)}{\int \tilde \pi(x_0) \tilde \rho(0, x_0) \dd x_0}
\end{equation}
for specified nonnegative $\tilde\pi$.  
Naturally, in case $\pi$ is itself Gaussian an obvious choice is to take $\tilde\pi=\pi$. 

\subsection{Parameter updating}\label{subsec:parameter}
Choose a Markov kernel $q$ for proposing new value for $\th$. 

\begin{alg}\label{mainalg-parest} {\it Parameter updating step}.
Suppose the current iterate is $(\th, x_0, Z)$ and $X=\GP_\theta(x_0, Z)$.
\begin{enumerate}
  \item Propose $\th^\circ$ from the proposal kernel $q$.
  \item If $(H, F, c)$ (defining the guiding term) depend on the parameter, then recompute the guiding term with parameter $\th^\circ$ using step 1 of Algorithm \ref{mainalg}. 
  \item Compute the corresponding guided proposal $X^\circ = \GP_{\theta^\circ}(x_0,Z)$. 
\item  Compute 
\[ A = \frac{q(\theta\mid \th^\circ)}{q(\theta^\circ \mid \theta)} \frac{\kappa(\theta^\circ)\tilde\rho_{\theta^\circ}(0,x_0)}{\kappa(\theta)\tilde\rho_\th(0,x_0)} \frac{\Psi_{\theta^\circ}(X^\circ)}{\Psi_\th(X)} \prod_{i=0}^{n} \frac{C_{i,{\theta^\circ}}(X^\circ_{t_i})}{C_{i,\th}(X_{t_i})} .
\]
 Draw $U\sim \scr{U}(0,1)$. If $U<A$ replace $X=X^\circ$ and  $\th=\th^\circ$.
\end{enumerate}
\end{alg}

The following Gaussian conjugacy result is sometimes useful; its proof is an elementary consequence of Girsanov's theorem.  
\begin{prop}\label{conjugacy}
Assume a diffusion $X$ with drift of the form
\[
b(t, x) = \phi_0(t, x) + \sum_{k=1}^K \theta_k \phi_k(t, x),
\]
where both the diffusion coefficient and the initial distribution of $X_0$ are independent of $\theta$. If the full path $(X_t, t \in [0,t_n])$ is observed and a priori $\theta \sim \N(0,\Gamma_0^{-1})$, then $\theta \mid X \sim \N( \Gamma^{-1}\mu, \Gamma^{-1})$ with 
\begin{align*}
\mu &= \int_0^{t_n} \Phi(t, X_t)'(\dd X_t - \phi_0(t, X_t)\dd t)\\
\Gamma &= \int_0^{t_n} \Phi(t, X_t)' a^{-1}(t, X_t) \Phi(t, X_t) \dd t + \Gamma_0,
\end{align*}
where  $\Phi(t, x) = (\phi_k(t,x))_{1 \le k \le K}$.
\end{prop}

\subsection{Initialisation}\label{subsec:init}

While any value of $(\th, x_0)$ within the support of their prior is possible, we propose to choose an initial value $\th$ and subsequently sample $x_0 \sim \pi^\circ$ for specified nonnegative $\tilde\pi$.  
Next, we sample a Wiener process $Z$ on $[0,t_n]$.
The corresponding guided proposal is obtained from   $X^\circ=\GP_\theta(x_0,Z)$, which can be computed using
\[ \dd X^\circ_t = b(t,X^\circ_t)\dd t + a(t,X^\circ_t)\left(F(t)-H(t)X^\circ_t\right) \dd t + \si(t,X^\circ_t) \dd Z_t. \]
Next, we initialise $X$ by defining $X=(X^\circ_t,\, t\in [0,t_n])$.

\subsection{Blocking strategy}\label{sec:blocking_strategy}
In this article, for comparisons of the effects of blocking we always adopt a ``chequerboard'' updating strategy. More precisely, we say that we use ``blocks of length $k$'' (for an even $k$) to mean that the following strategy is adopted for updating the path:

\begin{enumerate}
\item Initialise $X_{(0:n)}$.
\item For $i=1,\ldots, \lfloor n/k\rfloor$, sample bridges $X_{(ki-k: ki)}$, conditional on $X_{ki-k}$, $X_{ki}$ and $\{V_{ki-j};j=1,\dots , k-1\}$.
\item Sample the last segment $X_{(\lfloor n/k\rfloor k:n)}$ conditional on $X_{\lfloor n/k\rfloor k}$ and $\{V_{j};j=\lfloor n/k\rfloor k,\dots, n\}$.
\item For $i=1,\ldots, \lfloor n/k\rfloor-1$, sample bridges $X_{(ki-k/2: ki+k/2)}$, conditional on $X_{ki-k/2}$, $X_{ki+ki/2}$ and $\{V_{ki+j};j=-k/2+1,\dots, k/2-1\}$.
\item Sample the first segment $X_{(0:k/2)}$ conditional on $X_0$, $X_{k/2}$ and $\{V_{j};j=1,\dots, k/2\}$.
\item Sample the last segment $X_{(\lfloor n/k\rfloor k-k/2:n)}$ conditional on $X_{\lfloor n/k\rfloor k-k/2}$ and $\{V_{j};j=\lfloor n/k\rfloor k-k/2+1,\dots ,n\}$.
\end{enumerate}

Naturally, the algorithm above describes only smoothing. In order to transform the above into an inference algorithm a parameter update step described in Algorithm \ref{mainalg-parest} may be introduced in between steps 3 and 4 or after step 6 or both.

%---------------------------
%%%%%%%%%%%%%%%%%%%%%%%%

\section{Continuous-discrete smoothing for high dimensional diffusion processes}\label{sec:high}

For high dimensional SDEs with $d \gg 1$, the computation of the  $d\times d$  backward filtered covariance $P(t)$ in \eqref{eq:def_rtilde_in_H_and_nu} becomes prohibitive in an implementation in dense matrix 
algebra (and likewise the equations for its inverse). 
Another numerical difficulty is the computation of 
the likelihood accounting for the difference in $a$ and $\tilde a$ in case these are unequal. In this section we do not consider the latter problem, and therefore, restrict our attention to the high dimensional situation where $a = \tilde a$. In this case the linear filtering step is the numerical bottleneck and not the sampling step. Specifically, if the process $X$ and the observations are of the form
\begin{equation}\label{DSPDE}
 \begin{aligned}
\dd  X_t &= (B(t)  X_t + F(t,  X_t)) \dd t + \sigma_t \dd  W_t, \quad  X_0 \sim p({x}_0), \\
Y_i &= L   X_{t_i} + \eta_i, \quad \eta_i \sim \N(0, \Sigma_i),
\end{aligned}
\end{equation}
then a natural choice is to take $\tilde B(t) = B(t)$ and $\tilde \sigma(t) = \sigma(t)$.

Even if $\tilde B$ and $\tilde \sigma$ are sparse, $\Hd(t)$ typically is not. But if those operators that describe the drift and dispersion of $\tilde X$ are \emph{local}, then one may be able to approximate  $\Hd$ by a sparse matrix. Instead of stating a formal definition we provide an illustrative example, which can be seen as a variant of the stochastic heat equation on a graph with drift. 
Consider the $d$-dimensional diffusion $(X_t)$ for which a state $x$ is understood as a discrete approximation to a function $f\colon [0,1]\to \RR$, so that in particular, a coordinate $X^{(i)}$ corresponds to an evaluation of $f$ in $\tfrac{i-1}{d-1} \in [0,1]$. More concretely, let $X$ be defined as a solution to
\begin{equation}\label{langevin}
\dd X_t  = -  \tfrac{\sigma^2}2  (\Lambda+ c I) X_t  \dd t  + \sigma \dd W_t, 
\end{equation}
with
 $\sigma, c > 0$ and
\begin{equation}\label{laplace1d}
\Lambda = {\scriptstyle \left[\begin{smallmatrix} \phantom{-}1 & -1 &  & & \\
			-1 &\phantom{-} 2 & -1 &  &\\
			    & \ddots & \ddots & \ddots &\\
			    & & -1 & \phantom{-}2 &  -1  \\
			    &&& -1 & \phantom{-}1
\end{smallmatrix}\right]},
\end{equation}
the graph Laplacian of the linear graph with $d$ vertices corresponding to the coordinates. $\Lambda$ acts as a discrete approximation to the second derivative. $\Lambda$ is a local operator in the sense of having a representation by a banded matrix (this corresponds to all interactions in the linear graph being limited to only those between the neighbours). As a result, matrix-vector operations (matvecs) $\Lambda x$ can be computed efficiently in $\mathcal{O}(d)$ time. $X$ is a Gauss--Markov process with stationary distribution $\N(0, (\Lambda + c I)^{-1})$ (as $ \Sigma = (\Lambda + c I)^{-1}$ is a solution to the continuous time Lyapunov equation $B \Sigma + \Sigma B' + C = 0$, where $B = -\sigma^2/2(\Lambda + c I)$). Thus, in the stationary regime,  $X(t)$ at a fixed time $t$ is a draw from a high-dimensional Gaussian law approximating a spatial Gaussian process on $[0,1]$ with continuous realisations of H\"older regularity just below $\frac12$. While $ (\Lambda + c I)^{-1}$ is not a sparse matrix, the entries decay exponentially moving away from the diagonal. By Proposition \ref{prop:lyap}, the same holds for $\Hd(t)$.
Good, sparse approximations are obtained by setting $ (\Lambda + c I)^{-1}$   or $\Hd(t)$ to zero outside of a fixed band. 

In contrast, the forward simulation and the likelihood evaluation happen in $\RR^d$ and are fast, as long as fast matvec operations with the operator $H$ are available. This implies that the actual bottleneck of our approach is
in solving a continuous-discrete linear filtering problem and we can rely on a long sequence of work in this direction.

From the various possibilities we consider but two: a sampling approach related to the ensemble Kalman filter and a sparsity enforcing solver for \eqref{eq:recP}. 

\subsection{Ensemble backward filter}

Instead of solving \eqref{eq:recP} one can exploit that for the linear backward process with a reversed time axis 
\[
\dd \bar X_s = (\bar B(s) \bar X_s + \bar \beta(s)) \dd s + \bar\sigma(s) \dd \bar W_s, \bar X_0 \sim \N(\nu_{t_{i+1}}, P_{t_{i+1}})
\]
for $T-s \in (t_i, t_{i+1})$ with $\bar B(s) = - \tilde B({T-s})$, $\bar \beta(s) = -\tilde \beta({T-s})$ and $\bar\sigma(s) = \tilde \sigma(T-s)$,
it holds
\[
\expec{\bar X_s} = \nu(T-s) \quad \text{and} \quad \operatorname{Var}(\bar X_s) = P(T-s).
\] 
Here, $ \dd \bar W_s$ can be taken to be an It\^o integral; as $\tilde\sigma$ does not depend on space, there is no need for an  It\^o correction. 

This can be used to obtain estimates of $\nu(s)$ and a low rank approximation of $P(s)$ by sampling $\bar X^{(k)}$ a number of $K$ times with $K \ll d$ and approximating by the empirical covariance $\Sigma = \frac{1}{K-1} \sum \bar X^{(k)}_{T-s} X^{(k)\prime}_{T-s}$ . 
To obtain a positive definite (and invertible) estimate of $P$, one employs localization and sets $\hat P = \Sigma \circ \Lambda$ where $\circ$ is the Hadamard (elementwise) product with a positive definite sparse localisation matrix $\Lambda$ chosen to ignore long range dependencies, see \cite{HoutekamerMitchell2001}. The observations are taken into account by way of the usual ensemble Kalman filter 
\[
\bar X^{(k)}_{T - t_i}  = X^{(k)}_{T-t_i+} + \hat P(t+) L_i^{{\T}}\left(L_i \hat P(t+)  L_i^{{\T}}+\Sigma\right)^{{-1}}(v_i-L_i  X^{(k)}_{T-t_i+}),
\]
see \cite{Evensen}.

\subsection{Forced sparsity}\label{sub:forcedsparsity}
Secondly, we consider sparsity enforcing backward integrators, where we set
\[
\Hd(t-h) = \operatorname{drop}_{\epsilon}\left(\Hd(t) + (-\tilde B(t) \Hd(t) - \Hd(t) \tilde B(t)' + \tilde a(t))h \right). 
\]
Here 
\[(\operatorname{drop}_{\epsilon} A)_{ij} = \begin{cases} A_{ij} & |A_{ij}| > \eps\\ 0 & \text{otherwise}\end{cases}.
\] 
If $\Hd$ is well-approximated by a sparse matrix -- as illustrated for the case \eqref{langevin} -- then $\operatorname{drop}_{\epsilon} P_h$ will be sparse. We explore this approach in section \ref{sub:meteosat}.

\begin{rem}
A particularly relevant case of \eqref{DSPDE} is a process $X$ that is obtained from a spatial discretisation of the stochastic partial differential equation 
\begin{align}
\dd \mathbb X_t &= (-\mathbf A \mathbb X_t + \mathbf F(t, \mathbb X_t)) \dd t + 
\boldsymbol \sigma_t \dd \mathbb W_t, \quad \mathbb X_0 = \mathbbm{x}_0,
\end{align}
taking solutions in a Hilbert space $\mathbb H$, 
where $\mathbf A$ is a densely defined, positive definite operator  and
$\mathbb W$ is a $\mathbf Q$-Wiener process with a covariance operator $\mathbf Q$ given by a positive definite trace class operator, see \cite{DaPratoZabczyk2014}.
\end{rem}

%-----------------
\section{Examples}\label{sec:examples}

{\sloppy In all examples, the differential equations for $H(t)$, $F(t)$ and $c(t)$ have been solved using the $7$th order Runge--Kutta solver (cf. \cite{VernerRK}). 
The guided proposal is obtained by solving its SDE using Euler discretisation.
The code is available in the folders \texttt{scripts/papers} of the package \texttt{BridgeSDEInference.jl}\footnote{\url{https://github.com/mmider/BridgeSDEInference.jl/tree/master/scripts/papers/cts_discr_smooth}} (\cite{BridgeSDEInference}) and 
\texttt{scripts} of the package \texttt{BridgeSPDE.jl}\footnote{\url{https://github.com/mschauer/BridgeSPDE.jl/blob/master/scripts/smoothing.jl}}. Both packages build upon \cite{Bridge},
for the programming language Julia (\cite{Bezansonetal2017}).
}

\subsection{Lorenz system}\label{subsec:lorenz}
The Lorenz system is notable for having chaotic solutions for certain parameter values and initial conditions. It is described by the SDE with the drift and dispersion coefficient given by 
\[ b(x)=\Bm \th_1(x_2-x_1) \\ \th_2 x_1-x_2-x_1 x_3 \\ x_1 x_2 - \th_3 x_3 \Em \qquad \text{and} \qquad \si=\si_0 I_{3 \times 3}. \]

\begin{figure}[h!]
\begin{minipage}[t]{0.5\textwidth}
\includegraphics[width=1.0\linewidth]{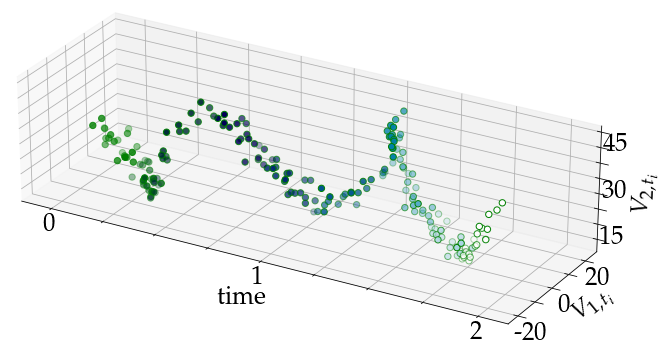}
\caption{Dataset 1}
  \label{fig:lorenz_data_dense}
\end{minipage}%
\begin{minipage}[t]{0.5\textwidth}
\includegraphics[width=1.0\linewidth]{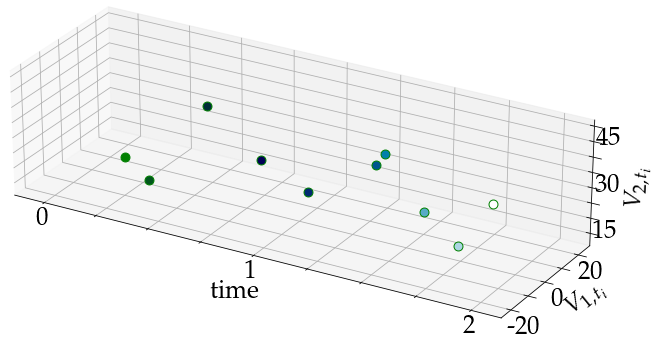}
\caption{Dataset 2}
  \label{fig:lorenz_data_sparse}
\end{minipage}
\end{figure}

{\bf Data:}
We initialised the process at $x_0=\Bm 1.5 & -1.5 & 25\Em^\T$ and simulated it on $[0,2]$ with mesh-width $2\text{e-}4$ and parameters $\th=\Bm 10&28&8/3\Em^\T$, $\si_0 =3$.
Only the 2nd and 3rd coordinate were observed so that  $L_i$ is a $2\times 3$ matrix with the first row $[0, \: 1, \: 0]$ and the second row $[0, \: 0, \: 1]$.
 We then defined two datasets.\\
{\bf Dataset 1:}
$200$ observations were retained at times $0.01, 0.02,\ldots, 2$, where we took
 $\Sigma_i=5\cdot I$ as the 
covariance matrix for the noise. \\
{\bf Dataset 2:}
$10$ observations were retained at times $0.2, 0.4,\ldots, 2$, where we took
$\Sigma_i=0.05\cdot I$.
The observed data are shown in figures \ref{fig:lorenz_data_dense} and \ref{fig:lorenz_data_sparse}. Note that the first coordinate process $\{X_{1,t};t\in[0,T]\}$ remains latent in both cases.

{\bf Algorithm details:}
Overall, six inference algorithms have been run, three on each dataset.
The procedures were inferring parameter $\th$, whereas parameter $\si_0$ was assumed to be known.
All four parameters could have been inferred simultaneously; however, by fixing $\si_0$ it was possible to illustrate the differences in performances of various algorithms more clearly.
Indeed, a varying $\si_0$ locally influences the speed at which chains on a path space, as well as parameter chains mix and this makes it more difficult to disentangle the contribution due to efficiency of the algorithm from that of a locally elevated or decreased volatility.
We illustrate the problem of inference for the volatility coefficient on a more challenging example in the following section.

We ran the following algorithms:
\begin{itemize}
    \item Dataset 1
    \begin{itemize}
        \item The algorithm of \cite{vdm-schauer}. It is a special case of the algorithm proposed in this article, in which a path---instead of being imputed in full---is updated in blocks of length $2$, as described in Section \ref{sec:blocking_strategy}.
        \item Inference algorithm with blocking, with blocks of length $8$.
        \item Inference algorithm with no blocking (where each proposal path is imputed in full).
    \end{itemize}
    \item Dataset 2
    \begin{itemize}
        \item Inference algorithm with no blocking.
        \item Inference algorithm with blocks of length 2.
        \item Inference algorithm with blocks of length 2 and adaptation of proposals laws.
    \end{itemize}
\end{itemize}
We set the following Gaussian priors over the parameter $\th$ and the value of the starting point:
\begin{equation*}
    \th\sim \N(0, \mbox{diag}(10^3,10^3,10^3)),\quad X_0\sim \N(x_0, \mbox{diag}(400,20,20)).
\end{equation*}
Setting a Gaussian prior over $\th$ made it possible to implement conjugate updates for $\th$ (cf. Proposition \ref{conjugacy} or \cite{vdMeulen}).
Each block had its own persistence parameter $\lambda_i$ of the preconditioned Crank--Nicolson scheme (in case of no-blocking a single parameter $\lambda$ was used).
For each block, $\lambda_i$ was tuned adaptively so as to target the acceptance probability of the imputation step for a given block to be $0.234$ (cf. Section 3 of \cite{RobertsRosenthal}).
The updates of the initial position were always done jointly with the updates of the path in the first block (or jointly with the entire path updates in case of no blocking).

We chose the auxiliary process according to the strategy \ref{enum:aux_linear_at_end}, where the initial guess for $\tilde{x}(t_i)$ was taken to be $\Bm \Xi_i & v_{i,1}& v_{i,2}\Em^\T$, $\Xi_i=25$, ($i=1,\dots,n$), and where the values of $\Xi_i$ were updated over the course of running an MCMC sampler: during the first $2500$ iterations, once in every $500$ steps we used an empirical average of $X_{1,t_i}$ to re-define $\Xi_i$. Note that this was done in all six experiments.

Additionally, in the experiment with the ``adaptation'', we used strategy \ref{enum:aux_linear_comb}, where the adaptation time was set to $2500$ iterations and adaptation updates were done once in every $500$ steps. The weights were changed from, initially, $(1,0)$ (the first value representing the weight for strategy \ref{enum:aux_linear_at_end}, the second one for strategy \ref{enum:aux_linear_at_mean}), through $(0.7, 0.3)$ and $(0.4, 0.6)$ up to $(0.2, 0.8)$ at iteration $1500$ (from then on the weights remained unchanged).

We set the density of the imputation grid to $2\cdot10^{-4}$.
The simulations took respectively: $784$s, $570$s, $198$s, $137$s, $260$s and $777$s to complete on an i5-5250U CPU, 1.60GHz.

\begin{figure}
\centering
\includegraphics[width=1.0\linewidth]{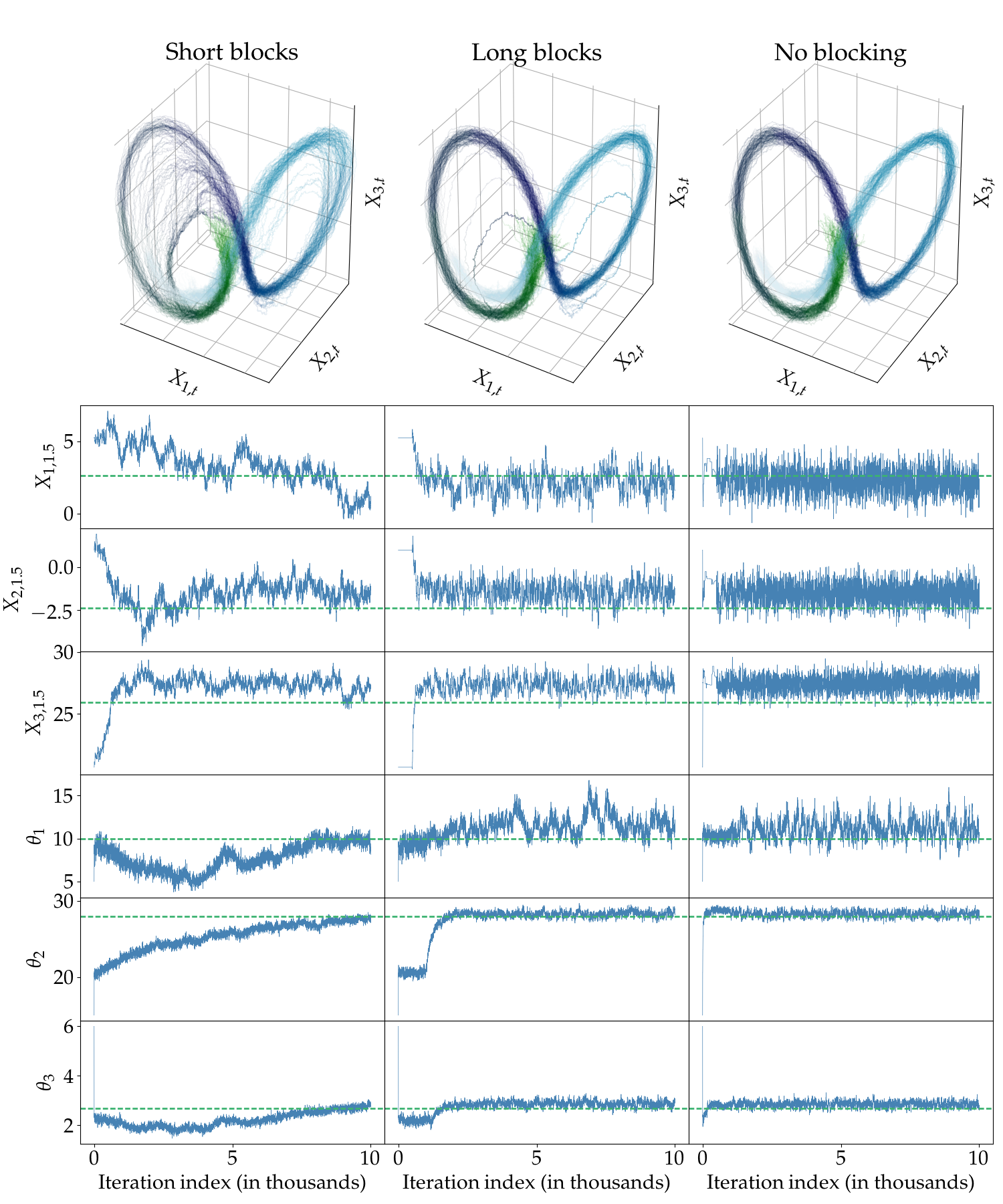}
\caption{Lorenz system: results on {\bf dataset 1}. Left column: results of a single MCMC run with blocks set to comprise of two inter-observation intervals (as in \cite{vdm-schauer}); middle column: blocks comprising of eight inter-observation intervals; right column: no blocking. Top plots: paths sampled by the MCMC samplers ($1$ in every $100$ sampled paths); Rows 2--4: marginal distribution of the coordinates of $X$ at time $t=1.5$. Rows 5--7: traceplots for the updated parameters $\th$.}
\label{fig:lorenz_system_dense_obs_results}
\end{figure}

{\bf Results:}
The results of the experiments associated with the first dataset are shown in Figure \ref{fig:lorenz_system_dense_obs_results} and they illustrate the outcome of only the first $10^4$ iterations. The summary concerning time-adjusted effective sample size that is based on the entire chain is given in Table \ref{tab:lorenz_experiment_dense}.

The green dashed lines in the top-most three traceplots indicate the values taken by the diffusion trajectory that was used to produce the data.
As we only have one realisation of the path and finitely many observations, the conditional  distribution (given the data) is not centred at this line; however, we can expect the samples to be close to them.
The green dashed lines in the bottom three traceplots indicate the values of the parameters that were set to simulate the data.

Under the first observational setting the discrepancy between an approximate and a true guiding term is small and thus blocking becomes a hindrance that decelerates mixing of the parameter chains and the path chain.
This is clearly illustrated in the left-most column of Figure \ref{fig:lorenz_system_dense_obs_results}, which corresponds to the algorithm of \cite{vdm-schauer}, where the block-length is set to $2$.
By increasing the block-length to $8$ (centre column) the mixing is improved and by removing it altogether and updating the entire path at once even better mixing can be achieved (right column).
The same conclusions are strongly supported by the results summarised in Table \ref{tab:lorenz_experiment_dense}, where the difference in the time-adjusted effective sample size is in excess of 2 orders of magnitude (we used a standard definition of the effective sample size, recommended by R. Neal in the panel discussion of \cite{kass1998markov}: $\texttt{ESS}(\theta):=n_{\texttt{ps}}/(1+2\sum_{k=1}^\infty\rho_k(\theta))$, with $n_{\texttt{ps}}$ denoting the number of posterior samples and $\rho_k$ the autocorrelation at lag $k$; it is implemented as a function $\texttt{ESS}$ in $\texttt{R}$ programming language).

In the top row of Figure \ref{fig:lorenz_system_dense_obs_results} we plotted a thinned chain of the imputed trajectories (1 in every 100 sampled paths is given; note that unlike figures \ref{fig:lorenz_data_dense} and \ref{fig:lorenz_data_sparse}, the temporal component is represented only by the changing colour of the trajectories, whereas each of the three axes have purely spatial meaning and correspond to the coordinates of the process $X$).
Notice how short blocks cause a slowdown in mixing on a path space, effectively disallowing the sampler to perform large moves. This is expected to be of a particular hindrance to sampling of multimodal diffusions.

The results from the first dataset clearly illustrate that removing the restriction on the block length can have a positive effect on the efficiency of sampling on a path space.
Nonetheless, some caution must be executed in extrapolating those conclusions.
To see this, consider the second dataset.
It is qualitatively different from the first one, in that the inter-observation distance is large enough for the non-linear dynamics of the target process to become pronounced between any two observations, whereas the decreased observational noise elevates the impact of the guiding term (for this reason it is also considered to be a substantially more difficult problem, for which a smaller number of numerical schemes can be applied to).

\begin{figure}
\centering
\includegraphics[width=1.0\linewidth]{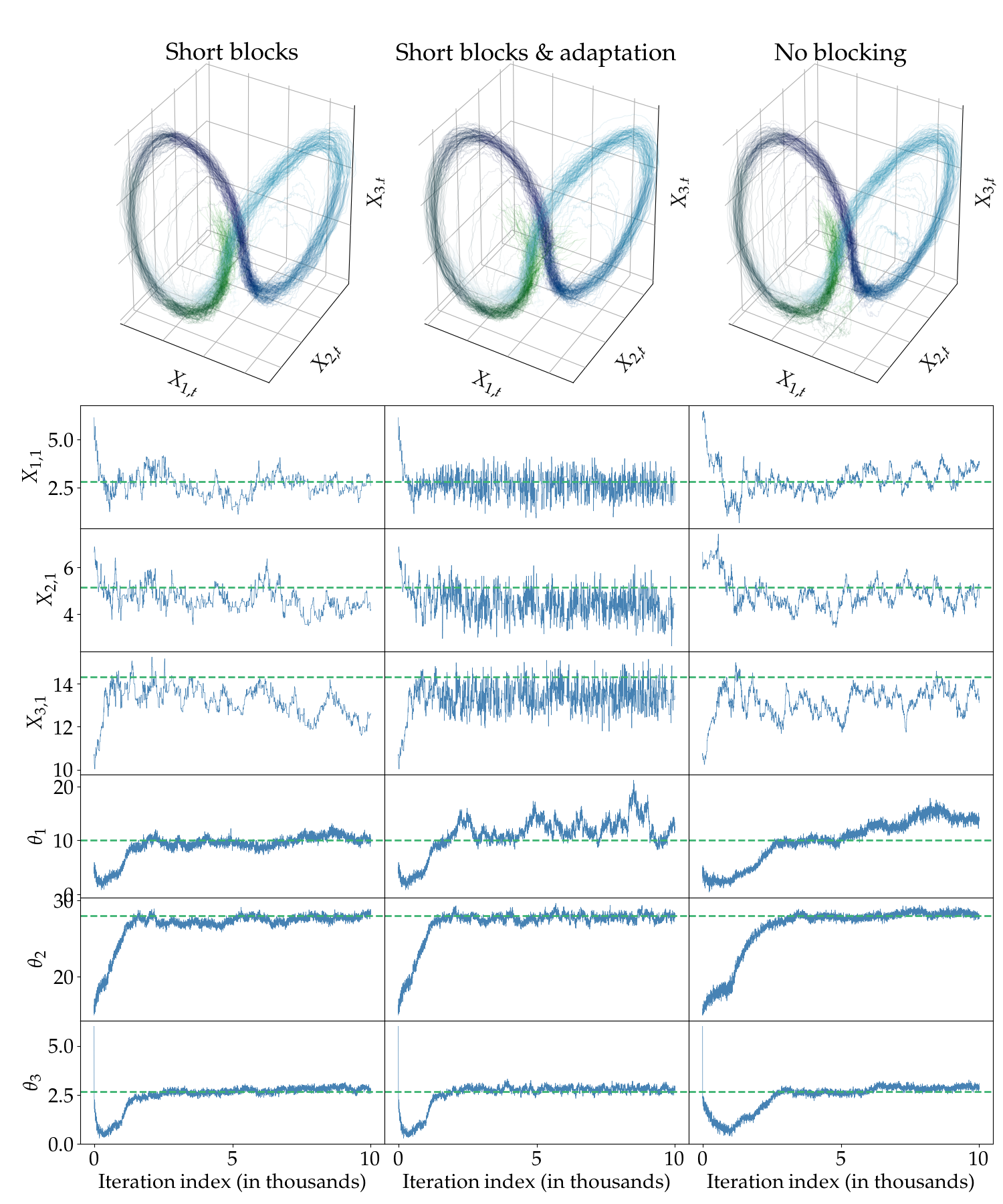}
\caption{Lorenz system: results on {\bf dataset 2}. Left column: results of a single MCMC run with no blocking; middle column: blocks comprising of two inter-observation intervals; right column: blocks of length 2 and adaptive tuning of the auxiliary law according to $b^{(aux)}=wb^{(7.1)}+(1-w)b^{(C)}$ (see Section \ref{sec:blocking_strategy} for details). Top plots: paths sampled by the MCMC samplers ($1$ in every $100$ sampled paths); Rows 2--4: marginal distribution of the coordinates of $X$ at time $t=1$. Rows 5--7: traceplots for the updated parameters $\th$.}
\label{fig:lorenz_system_sparse_obs_results}
\end{figure}

The inference results obtained on the second dataset are gathered in Figure \ref{fig:lorenz_system_sparse_obs_results}. In this setting, using blocking yields similar results as imputing the path in full. In fact, as summarised by Table \ref{tab:lorenz_experiment_sparse}, the values of taESS corresponding to parameter chains are approximately twice as large when blocking is used (as compared to a no-blocking scheme). However, it is possible to improve the mixing of the chains even further---as depicted in the centre column of Figure \ref{fig:lorenz_system_sparse_obs_results}---by designing better proposal laws.
We used a scheme from \ref{enum:aux_linear_comb} simply to motivate the usefulness of making the coefficients of the auxiliary diffusion time-dependent. The topic of optimal choice of the auxiliary law is beyond the scope of this paper.

Problems encountered in practice might fall anywhere on the spectrum spanned by the two datasets above or exhibit peculiarities of its own. We therefore believe that removing restriction on the block length---as it is done in this article---is a particularly practical improvement.

For reference we also ran the algorithm on the second dataset with the choice of the simplest strategy \ref{enum:aux_zero_drift} and reported the result in Table \ref{tab:lorenz_experiment_sparse}.
\tablestretch{1.3}
\begin{table}
\begin{tabular}{ lrrr } 
\toprule
\multicolumn{4}{c}{taESS (ESS)} \\
    & short blocks & long blocks & no blocking\\
 \midrule
 $X_{1,1.5}$ & 0.164 (128.6) & 1.991 (1134.9) & \textbf{52.932} (\textbf{10480.3}) \\ 
 $X_{2,1.5}$ & 0.447 (350.1) & 5.164 (2943.4) & \textbf{115.610} (\textbf{22890.5}) \\ 
 $X_{3,1.5}$ & 0.545 (427.0) & 1.468 \phantom{0}(836.6) & \textbf{121.568} (\textbf{24070.2})\\ 
\midrule
 $\theta_1$ & 0.039 (30.6) & 1.643 \phantom{0}(936.6) & \textbf{23.194} \phantom{0}(\textbf{4592.4})\\ 
 $\theta_2$ & 0.081 (63.7) & 0.366 \phantom{0}(208.6) & \textbf{77.675} (\textbf{15379.5})\\ 
 $\theta_3$ & 0.057 (44.5) & 3.108 (1771.7) & \textbf{55.141} (\textbf{10917.7})\\ 
\bottomrule
\end{tabular}
\caption{Time-adjusted effective sample size (taESS) and effective sample size (ESS) for the experiments with the first dataset of the Lorenz example. Based on chains of length $10^5$. Best results are written in bold.}
\label{tab:lorenz_experiment_dense}
\end{table}

\begin{table}
\begin{tabular}{ lrrrr } 
\toprule
\multicolumn{5}{c}{taESS (ESS)} \\
 & short blocks & short blocks \& adpt & no blocking & strategy \ref{enum:aux_zero_drift}\\
 \midrule
 $X_{1,1}$ & 1.328 (332.1) & \textbf{4.140} (\textbf{3216.6}) & 1.271 (175.3) & 1.17 (56.2)\\ 
 $X_{2,1}$ & 1.566 (391.4) & \textbf{5.162} (\textbf{4011.0}) & 1.618 (223.3) & 0.600 (28.8)\\ 
 $X_{3,1}$ & 1.366 (341.2) & \textbf{4.608} (\textbf{3580.4}) & 1.469 (202.7) & 0.346 (16.5)\\ % moritz: rounded 3580.39
 \midrule
 $\theta_1$ & 0.416 (103.9) & \textbf{0.602} (\textbf{467.5}) & 0.280 (38.6) & 0.330 (15.8)\\ 
 $\theta_2$ & \textbf{1.307} (326.8) & 0.963 (\textbf{748.5}) & 0.589 (81.2) & 0.901 (43.3)\\ 
 $\theta_3$ & \textbf{0.76} (190.6) & 0.563 (\textbf{437.2}) & 0.488 (67.3) & 0.196 (9.39)\\ 
\bottomrule
\end{tabular}
\caption{Time-adjusted effective sample size (taESS) and effective sample size (ESS) for the experiments with the second dataset of the Lorenz example. Additional comparison for the experiments with the auxiliary law chosen according to the strategy \ref{enum:aux_zero_drift}. Based on chains of length $10^5$. Best results are written in bold.}
\label{tab:lorenz_experiment_sparse}
\end{table}

\subsection{Prokaryotic auto-regulatory gene network}
We consider a simplified model of the auto-regulated protein transcription studied in \cite{Golightly1,golightly2011bayesian}. It is based on the stochastic model describing $\lambda$ repressor protein cI of phage $\lambda$ in Escherichia coli introduced in \cite{arkin1998stochastic}. The simplified model is characterised by eight biomolecular reactions: the reversible repression of transcription in which a protein dimer $\texttt{P}_2$ attaches to certain sites on the DNA:
\begin{equation*}
    \mathcal{R}_1\colon\, \texttt{DNA} + \texttt{P}_2\to \texttt{DNA}\cdot\texttt{P}_2,\qquad \mathcal{R}_2\colon\,  \texttt{DNA}\cdot\texttt{P}_2\to\texttt{DNA} + \texttt{P}_2;
\end{equation*}
the transcription of DNA into mRNA and the subsequent translation and protein folding:
\begin{equation*}
    \mathcal{R}_3\colon\, \texttt{DNA}\to \texttt{DNA} + \texttt{RNA},\qquad \mathcal{R}_4\colon\,  \texttt{RNA}\to\texttt{RNA} + \texttt{P};
\end{equation*}
the reversible protein dimerisation:
\begin{equation*}
    \mathcal{R}_5\colon\, 2\texttt{P}\to\texttt{P}_2,\qquad \mathcal{R}_6\colon\, \texttt{P}_2\to2\texttt{P};
\end{equation*}
and the mRNA and protein degradation:
\begin{equation*}
    \mathcal{R}_7\colon\, \texttt{RNA}\to\emptyset,\qquad \mathcal{R}_8\colon\, \texttt{P}\to\emptyset.
\end{equation*}
The total number of DNA strands is assumed to be fixed: $\texttt{DNA}+\texttt{DNA}\cdot\texttt{P}_2=K$. It is well known that such system admits a diffusion approximation (called the chemical Langevin equation) and for the given system above, an associated SDE is given by:
\begin{equation*}
    \dd X_t = S(\theta \circ h(X_t) )\dd t + S\odot \gamma(\theta\circ h(X_t))\dd W_t,\quad t\in[0,T],\quad X_0=x_0,
\end{equation*}
where $\circ\colon \RR^d\times\RR^d\to\RR^d$ is a component-wise multiplication:
\begin{equation*}
    (\mu\circ\nu)_i=\mu_i\nu_i,\quad i=1,\dots,d,
\end{equation*}
the operation $\odot\colon \RR^{d\times d'}\times\RR^{d'}\to\RR^{d\times d'}$ is defined via:
\begin{equation*}
    (M\odot\mu)_{i,j}=M_{i,j}\mu_{j},\quad i=1,\dots,d;\,j=1,\dots,d',
\end{equation*}
the function $\gamma\colon\RR^d\to\RR^d$ is a component-wise square root:
\begin{equation*}
    (\gamma(\mu))_i=\sqrt{\mu_i},\quad i=1,\dots,d,
\end{equation*}
$S$ is the stoichiometry matrix:
\begin{equation*}
    S=\Bm 0 & 0 & 1 & 0 & 0 & 0 & -1 & 0\\
        0 & 0 & 0 & 1 & -2 & 2 & 0 & -1\\
        -1 & 1 & 0 & 0 & 1 & -1 & 0 & 0 \\
        -1 & 1 & 0 & 0 & 0 & 0 & 0 & 0
        \Em,
\end{equation*}
and the function $h$ is given by:
\begin{equation*}
    h(x)=(x_3 x_4, K-x_4, x_4,x_1,x_2(x_2-1)/2, x_3, x_1, x_2)^\T.
\end{equation*}
For this SDE $X_t=(\texttt{RNA}_t, \texttt{P}_t, (\texttt{P}_2)_t, \texttt{DNA}_t)$ and $W$ is an $8$-dimensional Brownian motion. For more details about the model and how to derive chemical Langevin equations see \cite{Golightly1}.

We simulated three datasets using the Gillespie algorithm. In all three cases the process was started from $x_0=(8,8,8,5)$ and we set $K=10$.

\noindent {\bf Dataset 1:} First, we reproduced the observational scheme of \cite{golightly2011bayesian}, where noisy estimates of the total number of proteins that are not attached to any DNA strands are recorded at integer times:
\begin{equation}\label{eq:protein_obs}
    V_{i}=\Bm 0 & 1 & 2 & 0\Em X_i + \eta_i,\quad \eta_i\sim \N(0,2^2),\quad i=1,\dots,100.
\end{equation}

\noindent {\bf Dataset 2:} Next, in addition to observations from \eqref{eq:protein_obs}, we assumed that for each $8$ observations of protein counts it is possible to obtain one noisy count of \texttt{RNA}, i.e. that:
\begin{equation}\label{eq:protein_obs_RNA}
    V_{i}=\Bm 1 & 0 & 0 & 0 \\
               0 & 1 & 2 & 0\Em X_i + \eta_i,\quad \eta_i\sim N\left(0,\Bm 1^2 & 0 \\ 0 & 2^2 \Em\right),\quad i=8,16,\dots,96,
\end{equation}
and that for $i\in\{1,2,\dots,100\}\backslash \{8,16,\dots,96\}$, $V_i$ is given by \eqref{eq:protein_obs}.

\noindent {\bf Dataset 3:} Finally, we took the second dataset and permuted the noisy \texttt{RNA} counts randomly. More precisely, if $V_{i}^{(j)}$ denotes the $i^{th}$ observations from the $j^{th}$ dataset ($i=1,\dots,100$; $j=2,3$) then:
\begin{equation*}
V_{i}^{(3)}=
\begin{cases}
    V_{i}^{(2)},& i\in\{1,2,\dots,100\}\backslash \{8,16,\dots,96\},\\
    \Bm V_{1,\zeta(i)}^{(2)} & V_{2,i}^{(2)} \Em,& i\in\{8,16,\dots,96\},
\end{cases}
\end{equation*}
with $\zeta(i)$ denoting the $i$-th element of a random permutation $\zeta$ of $(8,16,\dots,96)$.

{\bf Algorithm details:} We ran the inference algorithm using guided proposals without blocking. Note that due to a rectangular shape of the volatility coefficient it is impossible to employ guided proposals with blocking and a non-zero persistence parameter $\lambda$ of the preconditioned Crank--Nicolson scheme. Consequently, one can employ the algorithm of \cite{vdm-schauer} only in a special case when the observations are made on a sufficiently dense time-grid, for which it is possible to set $\lambda=0$.

We set the prior distribution over the starting position to be $X_0\sim \N(x_0,I_{4\times 4})$ and, following \cite{golightly2011bayesian}, a prior distribution over the logarithm of each inferred parameter to be $\texttt{Unif}([-7,2])$. We remark that the priors chosen by \cite{golightly2011bayesian} that truncate the natural support of the parameters might be problematic, especially for the first dataset, as the data is not informative enough to narrow down the support of some of the posteriors, causing the parameter chains to hit against the boundaries of the priors. Instead, using an exponential prior over each $\theta_i$---this being the maximum entropy distribution of all continuous distributions supported on $[0,\infty)$---might be a more suitable choice. For the sake of fair comparison we stick with the choice made by \cite{golightly2011bayesian}.

The MCMC sampler started off with updating each log-transformed parameter separately using Metropolis--Hastings algorithm with random walk proposals and adaptation as described in Section 3 of \cite{RobertsRosenthal}. After the first $12\cdot 10^3$ steps of parameter updates we computed the empirical covariance of the parameter chains and switched to joint proposals of \cite{haario2001adaptive} (we used a modified version from Section 2 of \cite{RobertsRosenthal}). We kept updating the empirical covariance matrix once in every $100$ steps of the algorithm. We chose the auxiliary law according to the strategy \ref{enum:aux_linear_at_end}, where during the first $10^4$ iterations of the Markov chain we updated our guesses for the latent positions of the path at the observation times once in every $200$ iterations.

\begin{figure}
\centering
\includegraphics[width=1.0\linewidth]{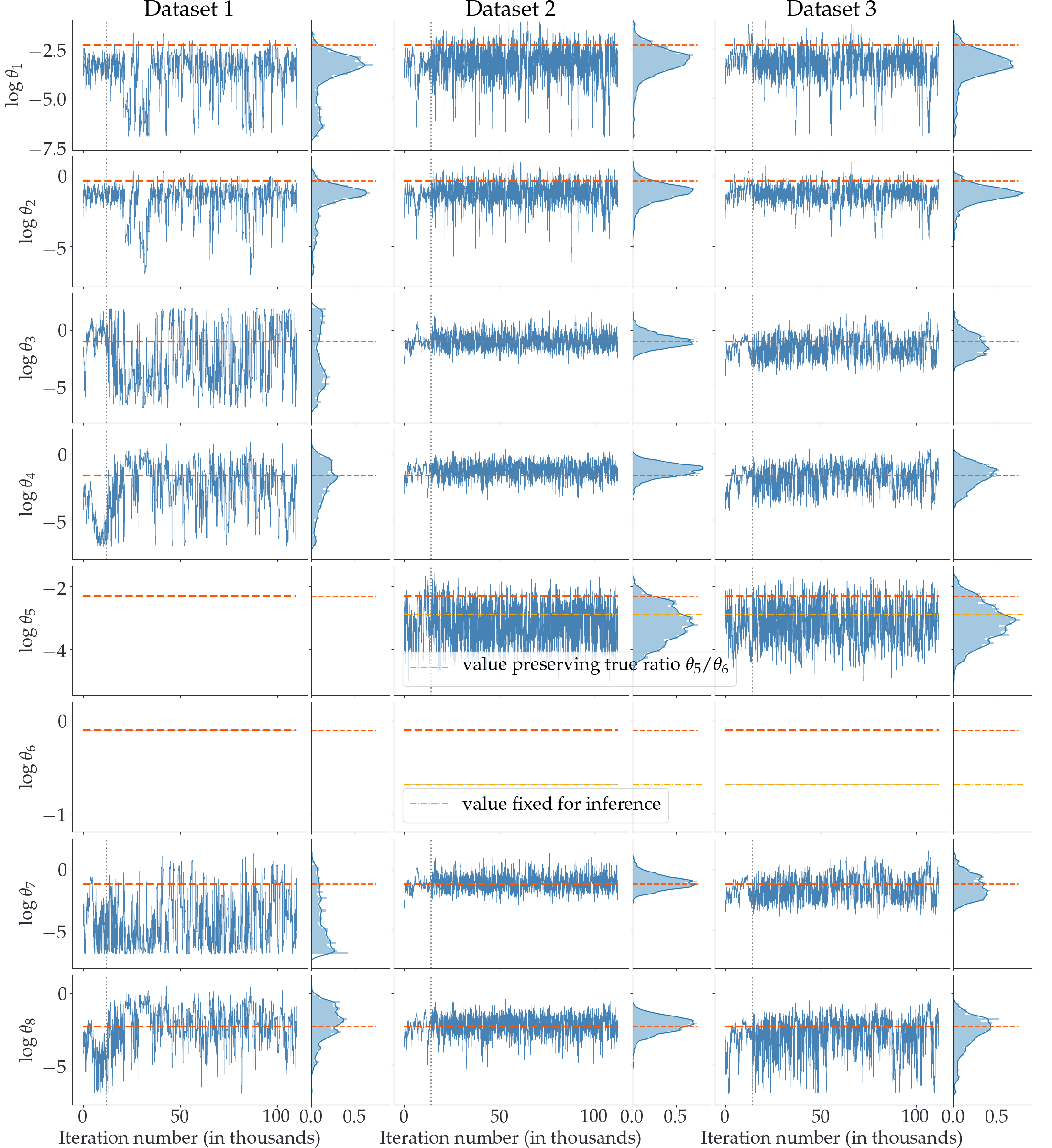}
\caption{Inference results for the prokaryotic example. Each iteration refers to a single step of parameter update. Left column gives the results of inference run on dataset 1; middle column: results from dataset 2; right column: results from dataset 3}
\label{fig:prokaryote_chains}
\end{figure}

{\bf Results:} Figure \ref{fig:prokaryote_chains} summarises the results from running inference on each of the three datasets. For each log-transformed parameter we give traceplots of the Markov chain and the resulting marginal density plots (with burn-in set to $2\cdot 10^4$ iterations). The true parameter values are marked with orange, dashed, horizontal lines, whereas grey, dotted, vertical lines mark the change in the type of the transition kernel used (from adaptive, single-site updates to joint updates of \cite{haario2001adaptive}). 

For the dataset 1 we fixed $\theta_5$, $\theta_6$ and $K$ to their true values and we aimed at inferring the remaining parameters. This is the same setup as in \cite{golightly2011bayesian}. The results are given in the left column of Figure \ref{fig:prokaryote_chains}.

Note that the marginal posteriors over $\log(c_3)$ and $\log(c_8)$ presented in \citet[Figure 3]{golightly2011bayesian} (corresponding to $\log(\theta_3)$ and $\log(\theta_8)$) concentrate decisively around incorrect values of the parameters, possibly pointing to some undiagnosed problems with the methodology. In our case, the true values of the parameters that were used to generate the data lie within the support of the posterior. In particular, the posteriors over $\log(\theta_3)$ and $\log(\theta_7)$ do not depart much from the priors, indicating that the data is not informative enough to identify those two parameters. Indeed, an examination of the reaction equations reveals that $\theta_3$ and $\theta_7$ regulate the production and the degradation of the \texttt{RNA} respectively and the observation scheme that provides only noisy counts of \texttt{P} might not provide sufficient information.

We generated the second dataset to examine the degree of improvement in identifying the true parameters that could be brought about by adding sparse observations of the \texttt{RNA}. We assumed that only $K$ is known and we fixed $\theta_6$ to an \emph{incorrect} but \emph{reasonable} value of $\theta_6^\star=0.5$. This was done to reproduce a real setting in which none of the $\theta$ parameters are known, so fixing any $\theta_i$ parameter to its true value is impossible. By fixing $\theta_6$ to an incorrect value and running the inference algorithm we hope to estimate $\theta_5/\theta_6$. Ideally, one would want to estimate all $8$ parameters simultaneously, but the dataset is not informative enough to conduct such inference. The results are summarised in the middle column of Figure \ref{fig:prokaryote_chains}. Parameters $\theta_i$, $i\in\{1,2,3,4,7,8\}$ have been successfully identified and the posterior over $\theta_5$ concentrates around the value $\theta_5^\star$ for which $\theta_5^\star/\theta_6^\star$ is equal to the true ratio $\theta_5/\theta_6$ of the parameters that were used to generate the data.

Naturally, the problem with the observational setting of dataset 2 is that noisy counts of $\texttt{RNA}_t$ need not be easily available in practice. However, it is conceivable that it is possible to collect information about the marginal distribution of \texttt{RNA}.  To make use of such information, at each time $t\in\{8,16,\dots,96\}$ instead of observing a true, noisy count of $\texttt{RNA}_t$ we could instead observe a realisation from the marginal density of \texttt{RNA}. We reproduce such observational setting by randomly permuting the time-labels of the generated $\texttt{RNA}_t$, $t\in\{8,16,\dots,96\}$, resulting in the dataset 3. The inference results in this setting (assuming $K$ known and $\theta_6$ fixed to an incorrect value of $\theta_6^\star=0.5$) are summarised in the right column of Figure \ref{fig:prokaryote_chains}. Some of the marginal posteriors appear flatter than the ones in the middle column, but the loss of information is not substantial and the posteriors still identify the true parameter values.

\subsection{Tracking the translation of a dynamic Gaussian random field}\label{sub:meteosat}

\begin{figure}
\begin{centering}
\includegraphics[width=0.24\linewidth]{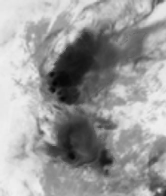}
\includegraphics[width=0.24\linewidth]{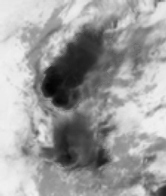}
\includegraphics[width=0.24\linewidth]{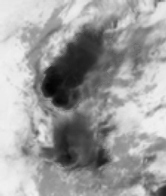}
\includegraphics[width=0.24\linewidth]{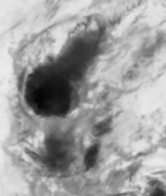}

\includegraphics[width=0.99\linewidth]{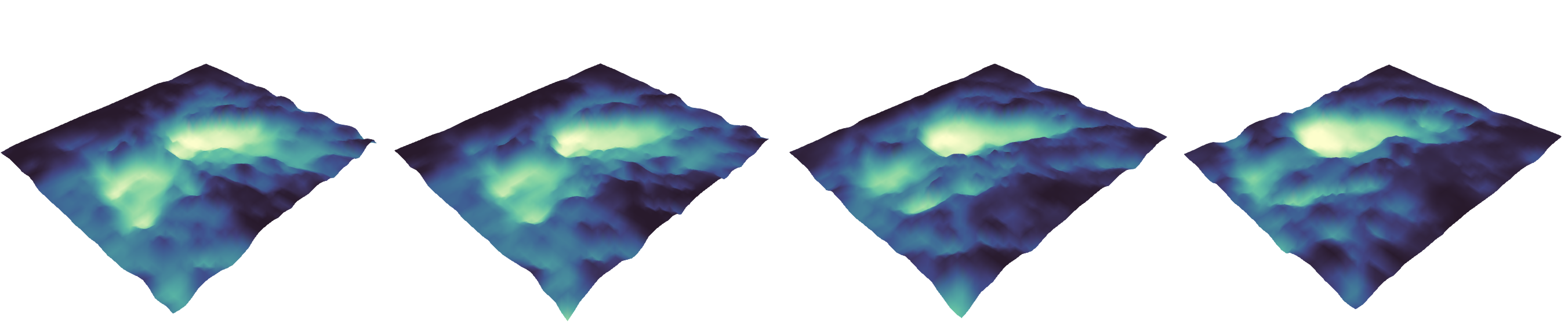}

\caption{Sequence of infrared Meteosat-MSG2 satellite images depicting
the evolution of a convective cloud system. Image from \citep{Avenel2014}.
}\label{meteosat}
\end{centering}
\end{figure}

We consider the problem of inferring a latent spatial process from noisy low frequency observations. As an example we take a few pictures from a sequence of satellite images depicting the evolution of a convective cloud system taken 15 minutes apart. The sequence was analysed by  \cite{Avenel2014} from the perspective of curve tracking.\footnote{The images were processed to remove annotated curves.} A visual inspection indicates that the system moves south--east and we are interested in estimating the speed of the system, as well as its location and extent at intermediate times.

No attempt is made to physically model a convective system. This example is merely meant
to serve as a starting point for applications in high dimensional settings.
The statistical analysis is made on the minimalistic assumptions that we observe a large scale phenomenon moving at constant speed through space, but the images show additional temporal and spatial high frequency phenomena we consider as random and it is the latent, large scale evolution that we are interested in. In this way we develop a generic model to track deformations and translations in time not bound to this particular meteorological example. Both Wiener and formal observation noise then capture all dynamics, observation errors and local physical phenomena which cannot be explained by our simple model. 

We set this numerical experiment up in the way of a 2-$d$ generalisation of \eqref{langevin}.
It is convenient to describe the model as a discretely observed, matrix-valued Langevin process (stochastic heat equation) on the graph with a stationary distribution that shows large scale variation in the distribution of its samples.This is similar in spirit to \cite{HartogvanZanten2019}, but we take a more dynamical view with discrete time observations and continuous latent dynamics and a drift term. Furthermore, via forward guiding our approach has a natural non-Gaussian extension. To avoid working with covariance tensors describing the uncertainty of a matrix valued process we move to the equivalent, vector-valued process, obtained through the application of the $\operatorname{vec}$ (vectorisation) operation.

The images have a single channel with values in $[0,1]$, so each pixel corresponds to one entry of a real $m\times n$ matrix, but we do not take the boundedness into account and model them as real-valued matrices.  The matrix-valued observations are denoted by  $Y_{0}$, \dots, $Y_{3}$ corresponding to pictures with indices $33, 35, 40, 45$. The chosen images are not equally spaced in time, but this poses no difficulty to our method. The last three pictures are taken $75$ minutes apart, the first two pictures are $30$ minutes apart. We consider $75\,$min as $1$ time unit.
The original image size was $196\times166$ and it was downsampled to $ 65\times55$.

We assume that 
\[
Y_i = X_{t_i} + \eta_i, \eta_i \sim \N(0, \Sigma),
\]
where $\eta$ is matrix-valued Gaussian noise with covariance tensor $\Sigma$ of the appropriate dimension (corresponding to the covariance matrix of $\operatorname{vec}(\eta)$). We took $\Sigma = \sigma^2_\epsilon I$

The latent process is a stochastic process $X = (X_t)$ as well taking values in the set of $m\times n$-real matrices, so nominally a 3575 dimensional diffusion process solving
\begin{align}\label{langevin2d}
  \dd X_t  &= -\tfrac{\sigma^2}2  (\rho\Lambda + cI) X_t \dd t  + F_\theta(X_t) \dd t + \sigma \dd W_t,
\end{align}
 with $ W_t$ an uncorrelated, matrix-valued Brownian motion, $\sigma, \rho, c > 0$ and
 $\Lambda$ the graph Laplacian tensor defined in \eqref{eq:graph_laplacian_tensor_meteo} below.
 
 The pixel indices $i,j$  are identified with the vertices  
$V = \{(i,j), i \in \{1, \dots, m\}, j \in \{1, \dots, n\}\}$  
of the $m\times n$-lattice $(V, E)$ with edges $E = \{ \{v, v'\}\colon v = (i,j), v'=(i',j') \in V, 
|i - i'| + |j - j'| = 1\}$ (using the set notation for edges). Thus, edges connect a pixel to its vertical and horizontal neighbours. Then 
  \begin{equation}\label{eq:graph_laplacian_tensor_meteo}
 \Lambda_{v, v'}
=
 \begin{cases}
\operatorname{degree}(v) & v = v'\\
-1 & \{v,  v'\} \in E\\
0 & \text{otherwise},
\end{cases}
\end{equation}
 which generalises \eqref{laplace1d},
and $F_\theta$, $\theta=(\theta_1, \theta_2)$ is a lateral translation of $\theta_1$ pixels south and $\theta_2$ pixels east. 
per unit of time, i.e.\
\[
F_\theta(x) = \theta_1 \rho \nabla^{\downarrow} x +  \theta_2 \rho \nabla^{\leftarrow} x,
\]
where $ \nabla^\downarrow = \nabla^{(0, -1)}$ and $\nabla^\leftarrow = \nabla^{(-1, 0)}$ are given by the mass transport operators
\[
\nabla^{(\Delta i, \Delta j)}_{v, v'}
=
 \begin{cases}
-1 & v  = v'\\
1 &  v = (i,j), v' = (i + \Delta i, j+\Delta j)\\
0 & \text{otherwise}
\end{cases}
\]
with $(\Delta i, \Delta j) \in \{(-1,0), (1,0), (0,1), (0,-1)\}$.
Thus $b_\theta(t, x) =  -\tfrac{\sigma^2}2  (\rho \Lambda + cI) x + F_\theta(x)$ with $\theta$ considered as unknown parameter is of the form prescribed by Proposition \ref{conjugacy}. Choosing $\rho$ and $\sigma$ proportional to image scale makes the model roughly scale invariant, which informally can be seen from considering the stationary distribution given by the Lyapunov equation under transformation  of the SDE  by a linear downsampling operator $R\colon \RR^{m\times n} \to  \RR^{ m/2   \times   n/2 }$, assuming $n$, $m$ are even. 
We then chose parameters which worked well at $20\times 24$ pixel resolution and scaled them to other resolutions  $41\times 49$ and $55\times 65$ that we use.
While the model works for finite dimensional problem at hand, the spatial white observation noise has no infinite dimensional scaling limit and for larger applications one would consider smoother observation noise with spatial correlation given by a non-diagonal covariance structure.

{\bf Parameter choices:} We set a scale $\rho	= 8m/195$ (which amounts to $8/3$ for the highest resolution level) and take $c = 0.1\rho$,  
We set  $\sigma = 0.08\rho$.  
 The noise level is set to $\sigma_\epsilon = 0.05\sqrt{\rho}$. We took the time step as $\delta t =  1/(35\rho)$.
We assume that $X_T$ is a priori Gaussian with mean 0.5 (a grey value) and variance $2.5I$ and equip $\theta$ with an improper uniform prior.

{\bf Algorithm details:} The SDE in our model is linear, so with $\tilde X = X$, the proposal is a sample from the conditional distribution and no change of measure is needed.
We solve the backward filtering equations from Theorem \ref{thm:noise-Hd} for $\nu$ and $\Hd$ using the Euler scheme in combination with the enforced sparsity method from Section \ref{sub:forcedsparsity} using Julia's native sparse matrix type (CSC). We took $\eps = 10^{-8}$.

We employed Rao-Blackwellisation to obtain an estimate of the mean of the joint posterior of parameters and latent path, where in one step, the mean of $X^\star$ given observations and $\theta$ is given by
\[ \dd x^\star_t = b_\theta(t,x^\star_t) \dd t+\sigma^2\Hd(t)^{-1}(x^\star_t - \nu(t)) \dd t, \quad x^\star_0 = \nu(0)
\]
and in the second step the mean of $\theta$ given $X^\star$ is obtained using Proposition \ref{conjugacy}.

For the correction formulas \eqref{eq:rec-Hd-init} we opted for an implementation based on the sparse Cholesky factorisation provided in Julia.  
\begin{figure}
%\hfill{}\includegraphics[width=0.24\linewidth]{meteo20-02-20/meteosat1}\hfill{}
%\opt{long}{
%\animategraphics[autoplay,loop,width=0.24\linewidth]{20}{meteo20-02-20/img15-}{1}{224}
%}\opt{short}{
%\animategraphics[autoplay,loop, every=5, width=0.24\linewidth]{10}{meteo20-02-20/img15-}{1}{224}
%}\opt{still}{
%\includegraphics[width=0.24\linewidth]{meteo20-02-20/img15-100}
%}
%\hfill{}\includegraphics[width=0.24\linewidth]{meteo20-02-20/meteosat4}\hfill{}
\includegraphics[width=0.3\linewidth]{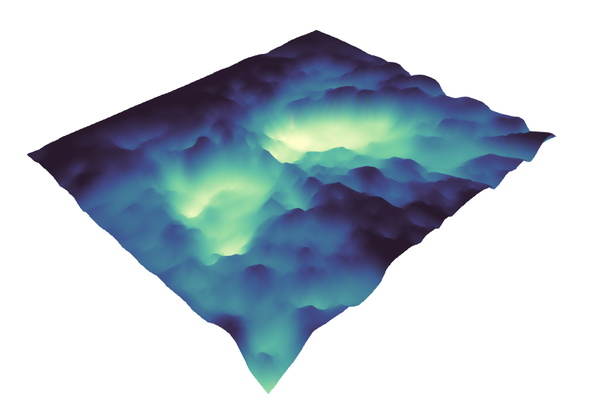}%
\opt{long}{%
\animategraphics[autoplay,loop,width=0.3\linewidth]{20}{meteo20-02-20/surf15-}{1}{224}%
}\opt{short}{%
\animategraphics[autoplay,loop,every=5,width=0.3\linewidth]{10}{meteo20-02-20/surf15-}{1}{124}%
}\opt{still}{%
\includegraphics[width=0.3\linewidth]{meteo20-02-20/surf15-100}%
}%
\includegraphics[width=0.3\linewidth]{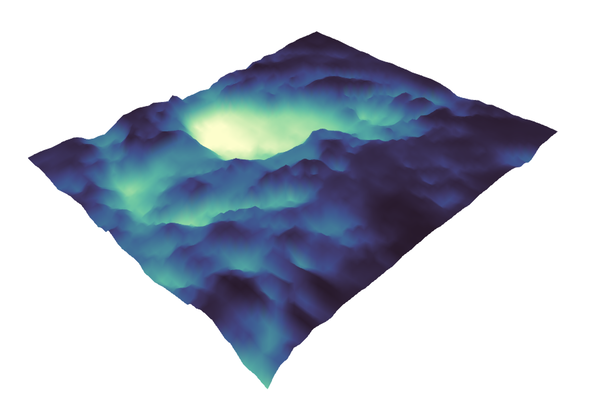}

\caption{In the \texttt{pdf} version: The center panel shows the posterior mean trajectory $x^\star$ as animation. The left frame shows the first observation and the right frame shows the last of the four observations in the same resolution, showing that the latent convective system bridge moves smoothly from initial to the final state. The animation is also available as supplementary file.}
\label{fig:latent-conv}
\end{figure}

\begin{figure}
\includegraphics[width=0.8\linewidth]{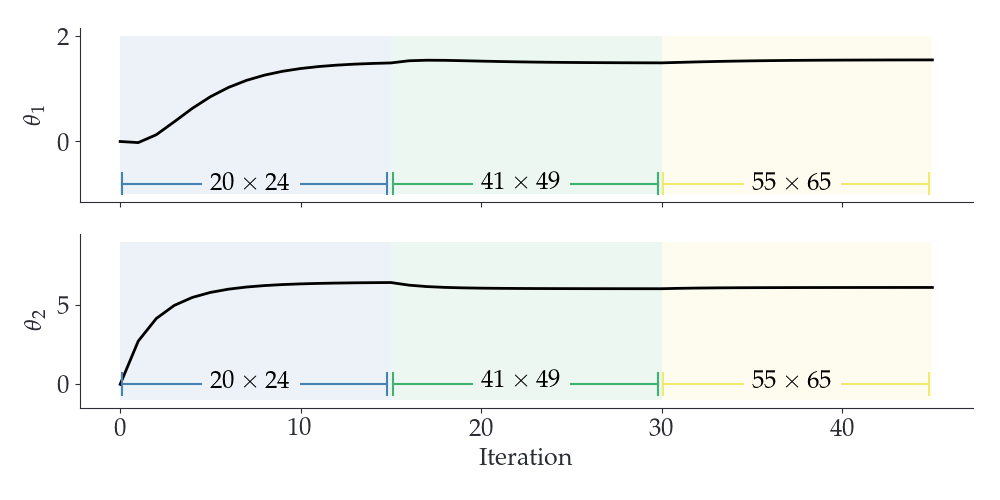}%../thetas}
\caption{Iterates of $\theta_1$, $\theta_2$ starting in $\theta_1 = \theta_2 = 0$ from Rao-Blackwellisation with increasing dimension.  First 15 iterates (blue): running with resolution $20 \times 24$. Next 15 iterates (green): continuing with resolution $41 \times 49$. Last 15 iterates (yellow): full resolution $55 \times 65$.}
\label{fig:raoblackwell}
\end{figure}

{\bf Results:} 
The first 15 steps of the algorithm were run on a model downscaled to lower resolution, for the subsequent 15 steps the resolution was increased and the final 15 steps were run on a model with the chosen, target resolution. Figure \ref{fig:raoblackwell} shows the trace of $\theta_1$, and $\theta_2$ for 45 iterates of the Rao-Blackwellised chain. After the last 15 steps the final estimates of the posterior means of the parameters are obtained as $\hat\theta_1 = 1.55$  and $\hat\theta_2 = 6.12$. These are in good agreement with visual assessment of the movement. The estimates of $\theta$ appear mostly stable at different model scales. (See the small changes in the estimates after each change of resolution at steps 15 and 30 in Figure \ref{fig:raoblackwell}.) The animation in Figure \ref{fig:latent-conv} shows the estimated posterior mean trajectory. The incorporation of discrete time observations with continuous latent dynamics and a drift term allows the algorithm to recognise the convective system as one object even if it has moved a substantial amount of pixels.

We only report rough timings: 
backward filtering step took half a minute for the correction steps and 
about two minutes for propagating the uncertainty at full resolution. Forward guiding took about 
half a minute and the parameter update about 1 second. (Timings are variable and depend on the current value of $\theta$, as $\theta$ influences the sparsity.) The Kalman gain had 3\,\% non-zero entries with  $\epsilon = 10^{-8}$.

%\newpage

%%%%%%%%%%%%%%%%%%%%%%%%%%%%%%%%%%%%

\appendix

\section{Related work}\label{sec:related}
If the full state of the diffusion process  is observed at all times $\{t_i,\, 0\le i \le n\}$ without noise, i.e.\ if $L_i=I$ and $\Sigma_i \equiv 0$ for all $i$ in \eqref{eq:lin-obs},  then the smoothing problem reduces to the sampling of $n$  independent diffusion bridges. This problem has attracted considerable attention over the past two decades, see for instance \cite{Eraker}, \cite{ElerianChibShephard}, \cite{DurhamGallant}, \cite{Clark}, \cite{Bladt2}, \cite{beskos-mcmc-methods}, \cite{HaiStuaVo09}, \cite{Schoenmakers}, \cite{LinChenMykland}), \cite{BeskosPapaspiliopoulosRobertsFearnhead}, \cite{DelyonHu}, \cite{lindstrom}, \cite{Schauer},  \cite{Whitaker} and \cite{Bierkens}. 
In the general case however, the connecting bridges cannot be sampled independently between adjacent observation times. In fact, at time $t\in (t_{i-1},t_i)$ the process $X$, conditioned on $\scr{V}_0$, depends on all future conditionings $V_i,\ldots, V_n$. To resolve this problem, subsequent simulation of bridges on overlapping intervals has been proposed by \cite{MR2422763}, \cite{Fuchs} and \cite{vdm-schauer}.

\cite{sarkka-sottinen} considered {\it filtering} (instead of smoothing) of diffusions  under the assumption that the dispersion coefficient is only allowed to depend on time. 
If the diffusion can be transformed to unit diffusion coefficient, then filtering can also be accomplished using the exact algorithm for simulation of diffusions, as introduced by \cite{BeskosRoberts}. This algorithm forms the basis for the methods presented in   \cite{fearnhead-partfilt-diff} and \cite{olsson}. 
Various solutions to  the filtering and smoothing problem are further discussed in  \cite{SARKKA2013500}. Key to the proposed algorithms therein is the assumption that the distribution of $X_t$, conditional on the data $\scr{V}_0$ can be approximated by the normal distribution. 

Recently, \cite{graham2019manifold} considered the same problem as considered here using a sampler based on constrained Hamiltonian dynamics. The basic idea is to consider the solution to the SDE as a forward map of Wiener-innovations that are constrained to a manifold implied by the observations.  Our approach is rather different and based on forward simulating paths that directly take all  observations  into account. 

%%%%%%%%%%%%%%%%%

\section{Technical background on guided proposals}\label{sec:recapguided} 

Previous work on guided proposals includes:
\begin{enumerate}
  \item[(A)] In \cite{Schauer} the basic idea was introduced in the setting of uniformly elliptic diffusions (i.e.\ the case where for each $y\in \RR^d$ there exists an $\eps>0$ such that $\|\si(t,x)' y\| \ge \eps \|y\|^2$), with one segment and a ``full'' observation at the time of conditioning. More precisely, the aim is to simulate a diffusion, conditioned on $X_0=x_0$ and $X_T=x_T$, with $0<T$. 
  \item[(B)] In  \cite{vdm-schauer} this was extended to the setting of two future conditionings. More precisely, in this paper it was shown how to define guided proposals to sample from a  diffusion starting in $X_0=x_0$ and conditioned on $v_S \sim \N(LX_S, \Sigma_S)$ and $X_T=x_T$, where  $0<S<T$.
  \item[(C)] In \cite{Bierkens} the work on \cite{Schauer} was generalised in a different direction by incorporating hypo-elliptic diffusions, observed partially and without noise. More precisely, here it is assumed that $X_0=x_0$ and the conditioning is given by $L X_T= v_T$.
\end{enumerate}

In all cases, the SDE for the conditioned process is of the form \eqref{eq:form-xstar}. The function $\rho$ is specific to the type of conditioning and depends on the unknown transition densities $p$ of the diffusion $X$.

The results of Theorem 1 in \cite{Schauer} and Theorem 2.14 in \cite{Bierkens} (and thus, the absolute continuity of the laws $\P^\star$ and $\P^\circ$ and the expression \eqref{eq:goal}) are proven under the following conditions: {\it (i)}  boundedness and smoothness of the drift coefficient $b$ and dispersion coefficient $\si$; {\it (ii)}  ``matching conditions on the diffusivity (and possibly drift)''. 
In setting (A), the matching condition is given by $\tilde{a}(T) = a(T,x_T)$ which can always be ensured. In setting (C) the matching conditions are more subtle and more complicated. The results in \cite{Bierkens} strongly suggest that for the diffusivity of the auxiliary process   the condition $L a(T,x_T) L^\T = L \tilde{a}(T) L^\T$ suffices. In case of hypo-ellipticity, additionally the drift of the auxiliary process,   $\tilde{b}(t,x) = B(t) x + \beta(t)$, needs to satisfy the equation  $L b(T,x_T) = L \tilde{b}(T,x_T)$. These conditions are particularly important in the noiseless limit of the discrete time (partial) observations.

%%%%%%%%%%
\section{Derivation of the guiding term in $X^\star$}\label{sec:doob_h_transf}
Here we explain the type of conditioning induced by the guiding term in $X^\star$. In particular, the specific form of $\rho$ in \eqref{eq:generalpull-p}.

Let  $p$ denote the transition density of a diffusion process $Y$ solving an SDE with drift $b$ and diffusivity $\sigma$ on the (canonical) filtered probability space $(\Omega, \scr{F}, \{\scr{F}_t\}, \PP)$ with $\Omega= C([0,t_n],\RR^d)$.  Assume the hierarchical model
\begin{equation}\label{eq:vj}
\begin{split}	
                v_{i} \mid \xi_{i} & \simind k_i(\xi_i; v_i) \qquad 1\le i \le n \\
				 \xi_{1}, \ldots, \xi_{n} &\sim \prod_{i=1}^n p(t_{i-1}, \xi_{i-1}; t_i, \xi_i)
\end{split}
\end{equation}
with $\xi_0$ assumed fixed (known). Without loss of generality we assume $t\in [0,t_1)$.
 We will assume $\xi_j \in \RR^d$, $L_j \in \RR^{m_j\times d}$ so that  $v_j \in \RR^{m_j}$. 
  Define $\bs{\xi}=(\xi_1,\ldots, \xi_n)$ and $\bs{v}=(v_1,\ldots, v_n)$. 
Let
\[ \pi_{t,y}(\bs{\xi}) =p(t,y; t_1,\xi_1) \prod_{j=1}^n p(t_j, \xi_j; t_{j+1}, \xi_{j+1})  \] and denote  the density of $(\bs{\xi}, \bs{v})$, conditional on $Y_t=y$, by
\[ \zeta_{t,y}(\bs{\xi}, \bs{v}) = \pi_{t,y}(\bs{\xi}) \prod_{j=1}^n k_j(\xi_j, v_j) . \]
We apply Doob's $h$-transform with 
\[ h(t,y) = \frac{\int \zeta_{t,y}(\bs{\xi}, \bs{v}) \dd \bs{\xi}}{\int \zeta_{0,\xi_0}(\bs{\xi},\bs{v}) \dd \bs{\xi}}. \]
Set $Z_t = h(t,Y_t)$. It is easily verified that for any $t\in [0,t_1)$, $Z_t$ is a martingale with mean $1$. Define a new probability measure $\mathbb{Q}$ on $\Omega$ by the change of measure $\dd \mathbb{Q}|_{\mathcal{F}_t}= Z_t \dd \mathbb{P}|_{\mathcal{F}_t}$, $\forall t\in[0,t_1)$. By Girsanov's theorem it follows that under $\mathbb{Q}$
\[ \dd Y_t = b(t,Y_t) \dd t + \sigma(t,Y_t) \dd \tilde{W}_t + a(t,Y_t)^\T \nabla \log h(t,Y_t) \dd t, \]
where $\tilde{W}$ is Brownian motion under $\mathbb{Q}$. 
We have 
\begin{align*}
	\mathbb{E}_\mathbb{Q} f(Y_t) &= \mathbb{E}_\mathbb{P}[Z_t f(Y_t)] = \mathbb{E}_\mathbb{P}[h(t,Y_t) f(Y_t)] = \int f(y) p(0,\xi_0; t,y) h(t,y) \dd y \\ 
	& = \int f(y) 	p(0,\xi_0; t,y) \frac{\int \zeta_{t,y}(\bs{\xi}, \bs{v}) \dd \bs{\xi}}{\int \zeta_{0,\xi_0}(\bs{\xi},\bs{v}) \dd \bs{\xi}} \dd y \\ & = 
	\int \left( \int f(y) \frac{p(0,\xi_0; t,y) \zeta_{(t,y)}(\bs{\xi}, \bs{v})}{\zeta_{(0,\xi_0)}(\bs{\xi}, \bs{v})} \dd y\right) \psi(\bs{\xi}\mid \bs{v}) \dd \bs{\xi}
\end{align*}
where we multiplied by $1=\zeta_{(0,\xi_0)}(\bs{\xi}, \bs{v})/\zeta_{(0,\xi_0)}(\bs{\xi}, \bs{v})$, used  Fubini to arrive at the final equality and   $\psi$ is defined by 
\[ \psi(\bs{\xi}\mid \bs{v}) = \frac{\zeta_{0,\xi_0}(\bs{\xi}, \bs{v})}{\int \zeta_{0,\xi_0}(\bs{\xi}, \bs{v}) \dd \bs{\xi}}.
 \]
Therefore, we have 	
\[ 	\mathbb{E}_\mathbb{Q} f(Y_t) = \int \mathbb{E}_\mathbb{P}[f(Y_t) \mid Y_{t_1}=\xi_1,\ldots, Y_{t_n}=\xi_n] \, \psi(\bs{\xi}\mid \bs{v})\dd \bs{\xi} .\]
This shows that under $\mathbb{Q}$, first $\bs{\xi}$ is sampled from the density $\psi$, and subsequently the process $Y$ is conditioned on the event $\{Y_{t_1}=\xi_1,\ldots, Y_{t_n}=\xi_n\}$. Note that  sampling from $\psi$ corresponds to sampling from the posterior of $\bs{\xi}$ in the  statistical model specified by \eqref{eq:vj}, with data $\bs{v}$. In this model, $\pi_{0,\xi_0}$ can be viewed as the prior density of $\bs{\xi}$. 

To connect to the main text, simply note that  $Y$ under the measure $\mathbb{P}$ is $X$, whereas $Y$ under the measure $\mathbb{Q}$ is $X^{\star}$ and that the $h$ transform is in our case $\rho$.

%%%%%%%%%%%%%%%%%%%%%%%%%%%%%%%%%

%%%%%%%%%%%%%%%%%%%%%%%%%%%%%%%%%

%%%%%%%%%%%%%%%%%

\section{Proofs of theorems in Section \ref{sec:backwardfiltering} and additional remarks}\label{sec:proofs}

In the proofs we will assume observation times $0<S<T$ and observations $V_S \sim \N(L_S X_S, \Sigma_S)$ and $V_T \sim \N(L_T X_T, \Sigma_T)$ with $L_S \in \RR^{m_S\times d}$ and $L_T \in \RR^{m_T\times d}$;  the extension to multiple future conditionings being trivial. 

%%%%%
\subsection{Proof of Theorem \ref{thm:rhotilde_using_MLmu}}
\label{proof:rhotilde_using_MLmu}
Define $\Phi$ to be the solution to $\dd \Phi(t) = B(t) \Phi(t) \dd t$ with $\Phi(0)=I$. Set $\Phi(t,s)=\Phi(t)\Phi(s)^{-1}$.
Define
\[ \Upsilon(t) =\begin{cases} \Bm \Sigma_S & 0_{m_S \times m_T} \\ 0_{m_T \times m_S} & \Sigma_T\Em & \qquad t\in (0,S] \\  \Sigma_T & \qquad t\in (S,T]\end{cases}, \]
and notice (upon differentiation) that
\begin{equation}
	\label{eq:Ltilde}
	L(t)=\begin{cases} \Bm L_S \Phi(S,t) \ind_{(0,S]}(t) \\ L_T \Phi(T,t)\Em & \qquad t\in (0,S] \\ L_T\Phi(T,t) & \qquad t\in (S,T]
\end{cases},
\end{equation}
\begin{equation}\label{eq:defMdagger} \Md(t)=\int_t^T L(\tau) \tilde{a}(\tau) L(\tau)^\T \dd \tau +\Upsilon(t),\end{equation}
and
\begin{equation}\label{eq:defmu}
  \mu(t)=\int_t^T L(\tau) \beta(\tau) \dd \tau, \qquad t\in [0,T],
\end{equation}
are the solutions to ODEs \eqref{eq:ode-tildeL}, \eqref{eq:ode-Mdagger} and \eqref{eq:ode-mu} respectively.
Note that $\mu(t) \in \RR^{m(t)}$ and that $\Upsilon(t)$, $L(t)$, $\Md(t)$ and $v(t)$ have similar structures when $t$ is either in $(0,S]$ or $(S,T]$. The relations in \eqref{eq:change-at-S} can be verified by evaluating $L(t)$ for both $t=S$ and $t=S+$ (and similarly for $\Md(t)$ and $\mu(t)$).

As we assume $\Sigma_T$ to be strictly positive definite, matrix $M(t)=\Md(t)^{-1}$ exists for  $t\in [0,T]$ (see however Remark \ref{rem:caseSigmaT=0} in case of no noise on the observations). 

The expression for $\tilde{r}$ for $t\in (0,S]$  follows by extending Lemma 2.5 in \cite{vdm-schauer} to the case where not necessarily $\Sigma_T=0$ and $L_T=I$. The expressions for $t\in (S,T]$ follow from equation (4.1) in \cite{vdm-schauer}. 

To derive $\tilde\rho$, note by \citet[Section 5]{Schauer} that 
$\tilde{X}$ is a Gaussian process with conditional mean $\mu_t(s,x)=\EE[\tilde{X}_t\mid \tilde{X}_s=x]$ and covariance $K_t(s)={\rm Cov}(\tilde{X}_s,\tilde{X}_t)$, where:
\begin{equation}\label{eq:mean_cov_gsn_gp}
    \mu_t(s,x)=\Phi(t,s)x + \int_s^t\Phi(s,\tau)\beta(\tau)\dd \tau,\qquad
    K_t(s)=\int_s^t\Phi(t,\tau)\tilde{a}(\tau)\Phi(t,\tau)^\T \dd \tau.
\end{equation}
Then, clearly, $(V_S^\T ,V_T^\T )^\T =((L_S\tilde{X}_S+\eta_S)^\T ,(L_T\tilde{X}_T+\eta_T)^\T )^\T $ is Gaussian as well and its mean (denoted with $\bar{v}(t)$) and covariance (denoted with $\bar{\Omega}(t)$) are given by:
    \begin{equation*}
        \bar{v}(t)=\left(
        \begin{matrix}
            L_S\mu_S(t,x)\\
            L_T\mu_T(t,x)
        \end{matrix}
        \right),\quad
            \bar{\Omega}(t)=\left(\begin{matrix}
            L_SK_{SS}L_S^\T +\Sigma_S & L_SK_{ST}L_T^\T \\
            L_TK_{TS}L_S^\T  & L_TK_{TT}L_T^\T +\Sigma_T
            \end{matrix}\right),
\end{equation*}
where $K_{\tau \nu}=K_{\tau\wedge \nu}(t)$, $\tau,\nu\in[t,T]$. Careful comparison of $\bar{v}(t)$ and $\bar{\Omega}(t)$ with the definitions \eqref{eq:Ltilde}, \eqref{eq:defMdagger} and \eqref{eq:defmu} reveals that
\begin{equation*}
    \bar{v}(t)=L(t)x + \mu(t),\qquad \bar{\Omega}(t)=M^{\dagger}(t),
\end{equation*}
and the expression for $\tilde\rho$ follows.

\begin{rem}
The value of  Theorem \ref{thm:rhotilde_using_MLmu} lies in recognition of the structure on both $H$ and $\tilde{r}$, something which was not noticed in \cite{vdm-schauer}. The theorem shows that both quantities can be written in a unified way on both $(0,S]$ and $(S,T]$. The key to this is the proper definition of $L$ (including the indicator). 
\end{rem}

\begin{rem}\label{rem:caseSigmaT=0}
Suppose $L_T=I$ and $\Sigma_T\equiv 0$. For $t\in (S,T]$,  $M(t)$ exists if and only if $\int_t^T \Phi(T,t) \tilde{a}(\tau) \Phi(T,t)^\T \dd \tau$ is invertible. This matrix is the controllability Grammian. Systems theory provides sufficient conditions for controllability. In case $\tilde{a}(t)$ is not invertible, then $M(t)$ exists if and only if the pair of functions $(B, \tilde\si)$ is 
controllable on $[t,T]$ for any $t\in [0,T)$ (cf.\ Section 5.6 in \cite{Karatzas-Shreve}).
\end{rem}

\begin{rem}
Suppose $L_S=I$ and $\Sigma_S=0$. This corresponds to the case where the diffusion is fully observed at time $S$ without noise. 	By the Markov property, the pulling term should only depend on $v_S$ (which is then in fact $x_S$) and not on $v_T$. To verify this, first note that we can write 
\[ \Md(t)=\Bm A & C \\ C^\T & D\Em, \]
where for $t\in [0,S)$
\begin{align*}
 A&= \int_t^S L_S\Phi(S,\tau) \tilde{a}(\tau) \Phi(S,\tau)^\T \dd \tau, \\	
 C&= \int_t^S L_S\Phi(S,\tau) \tilde{a}(\tau) \Phi(T,\tau)^\T L_T^\T \dd \tau= A \Phi(T,S)^\T L_T^\T, \\
 D&= \int_t^T L_T \Phi(T,\tau) \tilde{a}(\tau) \Phi(T,\tau)^\T L_T^\T \dd \tau. 		
\end{align*}
Now assume $L_S=I$ and that $A$ is invertible. Then we have 
\[ L(t)=\Bm L_S \\ L_T \Phi(T,S)\Em \Phi(S,t) = \Bm I \\ C^\T A^{-1}\Em \Phi(S,t). \]
Let $Z=(D-C^\T A^{-1} C)^{-1}$. Using the formula for the inverse of a block matrix, we get
\begin{align*}
	H(t)&= \Phi(S,t)^\T \Bm I& A^{-1} C\Em \Bm A^{-1} + A^{-1} C Z C^\T A^{-1} & -A^{-1} C Z \\ -ZC^\T A^{-1} & Z\Em \Bm I \\ C^\T A^{-1} \Em \Phi(S,t) \\
&=\Phi(S,t)^\T A^{-1} \Phi(S,t)= \left(\int_t^S \Phi(t,\tau) \tilde{a}(\tau) \Phi(t,\tau)^\T \dd \tau\right)^{-1}.
\end{align*} 
This is exactly as in Lemma 6 of \cite{Schauer}. 
\end{rem}

\begin{rem} 
Suppose  $t\in (t_{i-1},t_i]$ and we condition on future incomplete observations $(v_i,\ldots, v_n)$. In that setting the correct definitions for $\Upsilon$ and $L(t)$ are
\[ \Upsilon(t) =\diag(\Sigma_i,\ldots, \Sigma_n) \qquad \text{and}\qquad 
 L(t)=\Bm L_i \Phi(t_i,t) \\ 
L_{i+1} \Phi(t_{i+1},t) \\\cdots \\  L_n \Phi(t_n,t) \Em \].
\end{rem}

%%%%%%%%%%%%%%%%%%%%%
\subsection{Proof of Theorem \ref{thm:FHc_ODEs}}\label{proof:FHc_ODEs}

The expression for $\log\tilde\rho$ follows from Theorem \ref{thm:rhotilde_using_MLmu} by simple algebra, with
\begin{equation}\label{eq:decomp_c}
    \begin{split}
        c(t)&=\frac{1}{2}C^{(1)}_t + C^{(2)}_t,\quad\mbox{where}\\
        C^{(1)}_t=(v(t) - \mu(t))^\T &M(t)(v(t)-\mu(t))\quad\mbox{and}\quad C^{(2)}_t=\log\left[\left(2\pi\right)^{m(t)/2}\lvert \Md(t) \rvert^{1/2}\right].
    \end{split}
    \end{equation}

\noindent\textbf{\emph{Backward ordinary differential equations}}\quad We derive the ordinary differential equations solved by $H(t)$, $F(t)$ and $c(t)$ in a neighbourgood disjoint from $\{S,T\}$. We start with $H(t)$.
\begin{equation}\label{eq:proof_ode_for_H}
    \begin{split}
        \frac{\dd}{\dd t}H(t) &= \left(\frac{\dd}{\dd t} L(t)\right)^\T M(t)L(t) + L(t)^\T \left(\frac{\dd}{\dd t}M(t)\right)L(t) + L(t)^\T M(t)\left(\frac{\dd}{\dd t} L(t)\right)\\
        &=-B(t)^\T L(t)^\T M(t)L(t) + L(t)^\T \left(\frac{\dd}{\dd t} M(t)\right)L(t) - L(t)^\T M(t)L(t)B(t)\\
        &\quad-B(t)^\T H(t) - H(t)B(t) + L(t)^\T \left(\frac{\dd}{\dd t}M(t)\right)L(t).
    \end{split}
\end{equation}
Since $M(t)=(M^{\dagger}(t))^{-1}$:
\begin{equation}\label{eq:proof_ode_for_H_2}
    \frac{\dd}{\dd t}M(t)=-M(t)\left(\frac{\dd}{\dd t} M^{\dagger}(t)\right)M(t) = M(t)L(t)\tilde{a}(t)L(t)^\T M(t),
\end{equation}
and thus:
\[
    L(t)^\T \left(\frac{\dd}{\dd t}M(t)\right)L(t)=H(t)\tilde{a}(t)H(t).
\]
Substituting this back into \eqref{eq:proof_ode_for_H} yields:
\begin{equation*}
    \dd H(t) = \left( -B(t)^\T H(t) - H(t)B(t) + H(t)\tilde{a}(t)H(t) \right)\dd t.
\end{equation*}
For $F(t)$ notice:
    \begin{equation*}
        \begin{split}
            \frac{\dd}{\dd t}F(t)&=\left(\frac{\dd}{\dd t} L(t)\right)^\T M(t)(v(t)-\mu(t))+ L(t)^\T \left(\frac{\dd}{\dd t} M(t)\right)( v(t)-\mu(t))\\
            &\qquad- L(t)^\T M(t)\left( \frac{\dd}{\dd t} \mu(t)\right) \\
            &=-B(t)^\T L(t)^\T M(t)\left( v(t)-\mu(t)\right)\\
            &\qquad+L(t)^\T M(t)L(t)\tilde{a}(t)L(t)^\T M(t)(v(t)-\mu(t))+L(t)^\T M(t)L(t)\beta(t)\\
            &=-B(t)^\T F(t)+H(t)\tilde{a}(t)F(t)+ H(t)\beta(t).
        \end{split}
    \end{equation*}
    For $c(t)$ we differentiate $C^{(1)}_t$ and $C^{(2)}_t$ in turn. By \eqref{eq:ode-mu}, \eqref{eq:proof_ode_for_H_2} and the definition of $F(t)$:
    \begin{equation}\label{eq:dC1}
    \begin{split}
      \frac{\dd}{\dd t} C^{(1)}_t &= -2\left(\frac{\dd}{\dd t}\mu(t)\right)^\T M(t)(v(t)-\mu(t))
      % \\&\quad
      + (v(t)-\mu(t))^\T \left(\frac{\dd}{\dd t}M(t)\right)(v(t)-\mu(t))\\
      &=2\beta(t)^\T L(t)^\T M(t)(v(t)-\mu(t))
      % \\&\quad
      +(v(t)-\mu(t))^\T M(t)L(t)\tilde{a}(t)L(t)^\T M(t)(v(t)-\mu(t))\\
        &=2\beta(t)^\T F(t) + F(t)^\T \tilde{a}(t)F(t).
    \end{split}
    \end{equation}
    To find $\frac{\dd}{\dd t} C^{(2)}_t$, first, note by \citet[eq. (10)]{minka2000old} that for a matrix-valued function $Y$:
    \[
      \frac{\partial \lvert Y \rvert}{\partial x} = \lvert Y \rvert\trace\left(Y^{-1}\frac{\partial Y}{\partial x}\right),
    \]
    which gives
    \begin{equation*}
      \begin{split}
        \frac{\dd\lvert\Md(t) \rvert}{\dd t} &= \lvert \Md(t) \rvert\trace\left(M(t)\frac{\dd \Md(t)}{\dd t}\right)\\
        &=\lvert \Md(t) \rvert\trace\left(  -M(t)L(t)\tilde a(t) L(t)^\T \right)\\
        &=-\lvert \Md(t) \rvert\trace\left(  L(t)^\T M(t)L(t)\tilde a(t) \right)=-\lvert \Md(t) \rvert\trace\left(  H(t)\tilde a(t) \right).
      \end{split}
    \end{equation*}
    Since at non-observation times $t$, $m(t)$ is constant, the chain rule implies
    \begin{equation}\label{eq:dC2}
      \frac{\dd}{\dd t} C^{(2)}_t = -\frac{1}{2}\trace(H(t)\tilde{a}(t)).
    \end{equation}
    Combining \eqref{eq:dC1} and \eqref{eq:dC2} yields the ODE for $c(t)$ stated in Theorem \ref{thm:FHc_ODEs}.

\noindent\textbf{\emph{Update equations}}\quad The boundary conditions for $H(T)$, $F(T)$, $C^{(1)}_T$ and $C^{(2)}_T$:
\begin{equation*}
    H(T)=L_T^\T\Sigma^{-1}_TL_T,\quad F(T)=L_T^\T\Sigma^{-1}_Tv_T,\quad C_T^{(1)}=v_T\Sigma^{-1}_Tv_T,\quad C^{(2)}_T=\frac{m_T}{2}\log(2\pi) + \frac{1}{2}\log \lvert\Sigma_T\rvert,
\end{equation*}
follow directly from the definitions: $L(T)=L_T$, $M(T)=\Sigma^{-1}_T$ and $\mu(T)=0$; as well as the expressions for $H(t)$, $F(t)$, $C^{(1)}_t$ and $C^{(2)}_t$ given in \eqref{eq:tildeH}, \eqref{eq:F} and \eqref{eq:decomp_c} respectively. The boundary condition for $c(T)$ follows immediately
\begin{equation*}
    c(T)=\frac{1}{2}C^{(1)}_T+C^{(2)}_T=\frac{1}{2}v_T\Sigma^{-1}_Tv_T+\frac{m_T}{2}\log(2\pi) + \frac{1}{2}\log |\Sigma_T|.
\end{equation*}
Now, combining the expressions for $H(t)$, $F(t)$ and $C^{(1)}_t$, given in \eqref{eq:tildeH}, \eqref{eq:F} and \eqref{eq:decomp_c} respectively, with the update equations for $L(S)$, $M(S)$ and $\mu(S)$ at time $S$, given in Theorem \ref{thm:rhotilde_using_MLmu}, yields the update equations for $H(S)$, $F(S)$ and $C^{(1)}_S$:
\begin{equation}\label{eq:updt_eq_atS_HFC1}
    H(S)=H({S+}) + L_S^\T\Sigma^{-1}_SL_S,\quad F(S)=F({S+})+L_S^\T\Sigma^{-1}_Sv_S,\quad C_S^{(1)}=C_{S+}^{(1)}+v_S\Sigma^{-1}_Sv_S.
\end{equation}
To derive the update equations for $C^{(2)}_S$, notice that at observation time $S$ we have
\begin{align*}
  \tilde\rho(S,x) &= \phi(v_S; L_S x, \Sigma_S) \tilde\rho(S+,x)\\
                  & = (2\pi)^{-m_S/2} |\Sigma_S|^{-1/2} \exp\left(-\frac12 (v_S-L_S x)^\T \Sigma_S^{-1} (v_S-L_S x)\right) \times \exp\left(-C^{(2)}_{S+}\right) \\
                  & \quad \times \exp\left(-\frac12 (v(S+)-\mu(S+)-L(S+)x)^\T M(S+)(v(S+)-\mu(S+)-L(S+)x)\right).
\end{align*}
This can be written as 
\[ \exp\left(-C^{(2)}_{S}\right) \exp\left( -\frac12 (v(S)-\mu(S)-L(S)x)^\T M(S) (v(S)-\mu(S)-L(S)x)\right), \]
upon defining
\begin{equation}\label{eq:rec_C2} C^{(2)}_S = C^{(2)}_{S+} + \frac{m_S}{2}\log(2\pi) + \frac{1}{2}\log |\Sigma_S| \end{equation}
and because we have
\[ v(S)=\Bm v_S \\ v(S+)\Em \quad \mu(S)=\Bm 0\\ \mu(S+)\Em \quad L(S)=\Bm L_S \\ L(S+)\Em \quad M(S)=\Bm \Sigma_S^{-1} & 0 \\ 0 & M(S+)\Em. \]
 Cobining \eqref{eq:updt_eq_atS_HFC1} and \eqref{eq:rec_C2} yields the update equation for $c(S)$
\begin{equation*}
    c(S) = \frac{1}{2}C^{(1)}_S + C^{(2)}_S = c({S+})+\frac{1}{2}v_S\Sigma_S^{-1}v_S + \frac{m_S}{2}\log(2\pi) + \frac{1}{2}\log|\Sigma_S|.
\end{equation*}

%%%%%%%%%
\subsection{Proof of Theorem \ref{thm:noise-Hd}}\label{proof:noise-Hd}

We first give the derivation for the ODE of $P(t)$. 
The derivation of the differential equation is the same whether we consider $t\in [0,S]$ or $t\in (S,T]$. In particular, the ODE for $H$ from Theorem \ref{thm:FHc_ODEs} and the definition of $\Hd$ imply that
\[ \frac{\dd \Hd(t)}{\dd t} = - \Hd(t) \left(\frac{\dd}{\dd t} H(t)\right) \Hd(t) =
\Hd(t) B(t)^\T + B(t) \Hd(t) -\tilde{a}(t).
\]
The update formula from time $S+$ to $S$ follows from the corresponding one in terms of $H$ and applying   Woodbury's formula.

The derivation of the differential equation for $\nu$ is the same whether we consider $t\in [0,S]$ or $t\in (S,T]$. %{eq:deriv-tildeM}
We have, using \eqref{eq:Hdrec}, \eqref{eq:proof_ode_for_H_2} and \eqref{eq:ode-tildeL}
\begin{equation}\label{eq:Hdrec} \frac{\dd}{\dd t}\left(\Hd(t) L(t)^\T M(t)\right)=B(t) \Hd(t) L(t)^\T M(t).\end{equation}
Using \eqref{eq:ode-mu} we get
\[ \frac{\dd}{\dd t}\left(v(t)-\mu(t)\right)=L(t)\beta(t). \]
The previous two equations together yield
\begin{align*} \frac{\dd}{\dd t} \nu(t) &= B(t) \Hd(t) L(t)^\T M(t)\left(v(t)-\mu(t)\right) + \Hd(t) L(t)^\T M(t)L(t)\beta(t) \\ & = B(t) \nu(t) + \beta(t). \end{align*}
The value of $\nu(T)$ follows from $\Hd(T)=(L_T^\T\Sigma_T^{-1}L_T)^{-1}$, $L(T)=L_T$, $M(T)=\Sigma_T^{-1}$, $v(T)=v_T$ and $\mu(T)=0$.

To obtain the expression for $\nu(S)$, note that 
\begin{align*}
\nu(S) & = \Hd(S) L_S^\T M(S) \left(v(S)-\mu(S)\right) \\ & 
= \Hd(S) \Bm L_S \\ L(S+)\Em \Bm \Sigma_S & 0 \\ 0 & M(S+)\Em^{-1}  \left(\Bm v_S \\v(S+)\Em -\Bm 0 \\ \int_s^T L_T \Phi(T,\tau) \beta(\tau)\dd \tau \Em \right) \\ & = \Hd(S) \left(L_S^\T \Sigma_S^{-1} v_S + L(S+) M(S+) \left(v(S+)-\mu(S+)\right)\right)	
\\ & = \Hd(S) \left(L_S^\T \Sigma_S^{-1} v_S +H(S+)\nu(S+)\right).
\end{align*}

\begin{proof}[Proof of Proposition \ref{prop:lyap}]
\begin{align*} \frac{\dd}{\dd t} P(t)&= \frac{\dd}{\dd t} \left(\Phi(t,T)(\Sigma+\Hd(T))\Phi(t,T) -  \Sigma \right) \\ &= \left(\frac{\dd }{\dd t}\Phi(t,T)\right)(\Sigma+\Hd(T))\Phi(t,T) 
  +\Phi(t,T)(\Sigma+\Hd(T)) \frac{\dd }{\dd t} \Phi(t,T)^\T \\ &
= B \Phi(t,T)(\Sigma + P(T)) \Phi(t,T)^\T  +  \Phi(t,T)(\Sigma + P(T)) \Phi(t,T)^\T B^\T\\ & 
= B (\Hd(t) + \Sigma) + (\Hd(t) + \Sigma) B^\T = B P(t) + P(t) B^\T - \tilde{a},
\end{align*}
where we used \eqref{eq:Pt} at the third equality and $\Sigma$ solving the Lyapunow equation at the final equality.
\end{proof}

\begin{rem}
We  investigate the behaviour of $\Hd(S)$ and $\nu(S)$ when the noise level tends to zero. {\it Assume    $L_S\Hd(S+) L_S^\T$ is invertible.} Then it follows from \eqref{eq:rec-Hd-init} that the expression for $\Hd(S)$ is also well defined when $\|\Sigma\| \to 0$.
Moreover, when $\Sigma=0$ we have
\[ L_S \Hd(S)  = L_S \Hd(S+) - L_S \Hd(S+) L_S^\T \left(L_S \Hd(S+) L_S^\T\right)^{-1} L_S\Hd(S+) = 0. \]

To evaluate the limiting behaviour of $\nu(S)$, we write 
\[	 L_S \nu(S) = L_S \Hd(S) L_S^\T \Sigma_S^{-1} v_S + L_S \Hd(S)  H(S+) \nu(S+). \]
The second term on the right-hand-side is easily seen to tend to zero when  $\|\Sigma_S\| \to 0$.
For deriving the limit of the first term on the right-hand-side we define $C=L_S \Hd(S+) L_S^\T \Sigma_S^{-1}$ and rewrite
\begin{align*} L_S\Hd(S) L_S^\T \Sigma_S^{-1} &= L_S\Hd(S+) L_S^\T \Sigma_S^{-1}\\ & \qquad -L_S\Hd(S+) L_S^\T \left(\Sigma_S + L_S\Hd(S+) L_S^\T \Sigma_S^{-1}\Sigma_S\right)^{-1} L_S \Hd(S+) L_S^\T \Sigma_S^{-1} \\ &= C-C (I+C)^{-1}  C = \left( I+C^{-1} \right)^{-1},
\end{align*}
where we used Woodbury's formula at the final equality. Now \[ C^{-1}= \Sigma_S \left(L_S\Hd(S+) L_S^\T\right)^{-1} \to 0,\quad \text{when}\quad  \|\Sigma_S\|\to 0.\] This implies that $ L_S\Hd(S) L_S^\T \Sigma_S^{-1}\to I$. 
Combining these results we obtain that 
\[ L_S \nu(S) \to v_S  ,\quad \text{when}\quad  \|\Sigma_S\|\to 0.\]

In particular, if we have full observations at time $S$, i.e.\ $L_S=I$, then  $\Hd(S)\to 0$ and $\nu(S) \to v_S$ if $\|\Sigma_S\|\to 0$.  As a consequence, in case of full observations and no noise, we recover the full observation without noise case (in this simpler setting, the differential equations for $\nu$ and $\Hd$ were derived in  \cite{vdm-schauer-residual}). 
However, in the general case  where $L_S$ is not of full rank, we need that $\det \Sigma_S \neq 0$ to obtain the value of $\nu(S)$ from $\nu(S+)$.

\end{rem}

%---------------

\section{Implementation using a time-change and scaling}\label{sec:timechange}

 If the noise level on the observation is small, care is required in the discretisation of guided proposals near the conditioning points. For this reason, a time-change and scaling was introduced in Section 5 of \cite{vdm-s-estpaper}. Here, we explain it for the case of 2 future conditionings. 
Define the mapping $\tau\,:\,[0,S+T] \to [0,S+T]$ by 
\[ \tau(s)=\begin{cases}
s\left(2-\frac{s}{S}\right) 	& \quad \text{if}\quad \quad s\in  [0,S]\\ 
 S+(s-S)\left(2-\frac{s-S}{T-S}\right) & \quad \text{if}\quad \quad s\in  [S,T] 
\end{cases}.
\]
For a mapping $f:\RR\to \RR$ we write $f_\tau(s)=f(\tau(s))$. 
We propose to discretise the process $U_s$, defined by 
\[ U_s =\frac{\nu_\tau(s)-X^\circ_{\tau(s)}}{\dot\tau(s)}, \]
instead on $X^\circ_s$. This process satisfies the SDE
\begin{align*}
\dd U_s =& \left( B_\tau(s)v_\tau(s)+\beta_\tau(s)-b_\tau(s,U_s))\right) \dd s \\ & - \left(\frac{\ddot\tau(s)}{\dot\tau(s)} I + a_\tau(s,U_s)J(s)\right)U_s \dd s\\& + \frac1{\sqrt{\dot\tau(s)}} \sigma_\tau(s,U_s) \dd W_s,\qquad U_0 =\frac{\nu_\tau(0)-x_0}{2}
\end{align*}
Here, $J$ is defined by 
\[ J(s) =\dot\tau(s) H_\tau(s).\] Furthermore, we have used the notation $b_\tau(s,y)=b(\tau(s),v_\tau(s)-\dot\tau(s) y)$ and similarly for $\si_\tau$ and $a_\tau$. 
For the likelihood ratio \eqref{eq:goal}, note that $\tilde{r}(\tau(s),X^\circ_{\tau(s)})=J(s) U_s$. 
\begin{align*}
 \int_0^T G(s,X^\circ_s) \dd s =& \int_0^T \left(b_\tau(s,U_s)-B_\tau(s,U_s)\right) J(s) U_s \dot\tau(s)  \dd s\\ &- \frac12 \int_0^T \mbox{tr}\,\left( \left\{a_\tau(s,U_s)-\tilde{a}_\tau(s)\right\} \left\{ J(s)-J(s) U_s U_s^\T J(s) \dot\tau(s)\right\} \right)\dd s
\end{align*}
Finally, by the chain-rule, $\Hd_\tau(s)$ and $\nu_\tau(s)$ satisfy the backward differential equations 
\[ \frac{\dd}{\dd t} \Hd_\tau(t)=\left( B_\tau(t) \Hd_\tau(t) + \Hd_\tau B_\tau(t)^\T - \tilde{a}_\tau(t)\right) \dot\tau(s) \] 
and
\[ \frac{\dd}{\dd t} \nu_\tau(t)= \left(B_\tau(t) \nu_\tau(t) + \beta_\tau(t)\right) \dot\tau(s). \]

\section*{Acknowledgments}
The research leading to the results in this paper has received funding from the European Research Council under ERC Grant Agreement 320637. M.M. was sponsored by the Max Planck Institute for Mathematics in the Sciences, Leipzig and the EPSRC [grant number EP/L016710/1].

%%%%
\bibliographystyle{harry}
%\bibliography{lit}
\bibliography{lit2}

\end{document}